# Military Applications of Machine Learning: A Bibliometric Perspective

**José Javier Galán [1,*], Ramón Alberto Carrasco [2] and Antonio LaTorre [3]**

1 Faculty of Statistics, Complutense University, Puerta de Hierro, 3728040 Madrid, Spain
2 Department of Management and Marketing, Faculty of Commerce and Tourism Complutense, University of Madrid, 28223 Madrid, Spain; ramoncar@ucm.es
3 Center for Computational Simulation (CCS), Universidad Politécnica de Madrid, 28660 Madrid, Spain; a.latorre@upm.es
* Correspondence: josejgal@ucm.es

**Abstract:** The military environment generates a large amount of data of great importance, which makes necessary the use of machine learning for its processing. Its ability to learn and predict possible scenarios by analyzing the huge volume of information generated provides automatic learning and decision support. This paper aims to present a model of a machine learning architecture applied to a military organization, carried out and supported by a bibliometric study applied to an architecture model of a nonmilitary organization. For this purpose, a bibliometric analysis up to the year 2021 was carried out, making a strategic diagram and interpreting the results. The information used has been extracted from one of the main databases widely accepted by the scientific community, ISI WoS. No direct military sources were used. This work is divided into five parts: the study of previous research related to machine learning in the military world; the explanation of our research methodology using the SciMat, Excel and VosViewer tools; the use of this methodology based on data mining, preprocessing, cluster normalization, a strategic diagram and the analysis of its results to investigate machine learning in the military context; based on these results, a conceptual architecture of the practical use of ML in the military context is drawn up; and, finally, we present the conclusions, where we will see the most important areas and the latest advances in machine learning applied, in this case, to a military environment, to analyze a large set of data, providing utility, machine learning and decision support.

## 1. Introduction

Machine learning (ML) allows the automation of many tasks by taking advantage of the large amount of information available from different sources, including big data applications. Its use is currently widely spread, and ML has become an important part of our daily lives [1].

In the military, the use of intelligent applications has also accelerated [2]. For example, the South Korean Ministry of National Defense has increased its information significantly, and with fewer and fewer intelligence analysts they need to apply artificial intelligence (AI) technology to process all the information in an accurate and timely manner [3]. Another example to note is the dependence on oil by military equipment and machinery. This is also where ML comes in, as military logistics must be intelligently based on informed deductions [4]; thus, we see how ML is integrated into the military world.

The objective of this paper is to present an architectural model that reflects how ML is applied in a practical way in the military environment. In this architecture, we solve aspects such as the most frequent data, algorithms and applications used in the military context.

While carrying out this work, as we will see in Section 2, we study related work, noting that there are few review works in this emerging topic, which has aroused our interest in performing a bibliometric analysis on one of the main scientific databases, Web of Science, up to and including the year 2021. In the same Section, we also present a conceptual architecture for the application of ML in a practical way in a nonmilitary organization, since there are no works reflecting such an architecture in the military domain.

The bibliometric methodology used in this work is explained in Section 3, and we will mainly make use of the SciMat bibliometric analysis tool, capable of performing a scientific mapping analysis in a longitudinal framework [5]. With this analysis we build a strategic diagram in which we identify the main areas of ML applied to the military field.

In Section 4, we apply the described methodology to perform an analysis according to origin: we see the main scientific areas in which ML is applied to the military world; author and citation: we determine who are the most active authors in this subject; country: we analyze how the countries that generate more scientific documentation in this sense are usually the ones that have fewer citations; and we distinguish two periods: before 2015 and after 2016, after which the increase in publications on ML in the military world rapidly ascends.

In Section 5, once this bibliometric analysis is completed, we are now in a position to redefine the conceptual architecture presented in Section 2 specifically for military organizations.

Finally, we come to some conclusions in which we expose the results obtained related to the main thematic areas found and the conclusions.

## 2. Related Work

First, in Section 2.1, we searched for bibliometric or review articles related to the application of ML in the military world, and we did not find relevant information. Then, in Section 2.2, we searched for a data-driven architecture for nonmilitary organizations to serve as a basis for establishing a new model oriented to the military world, which we complement with the present bibliometric study.

### *2.1. Previous Research on the Military Applications of ML*

In this Section, we studied review papers or bibliometric studies on AI (which includes ML) and related areas, in addition to their applications in the military field. The results of this study are shown in Table 1.

**Table 1.** Related review work.

| Category of the Review Work | Refs. |
|---|---|
| Robotics and smart devices with military applications | [6–22] |
| Generic ML and optimization techniques with military applications | [23–33] |
| ML and optimization techniques focused on military applications | [34–36] |

We have classified these works into three categories:

- Reviews on robotics and smart devices with military applications. We found several works on this subject, including on drones, sensors, computer vision, unmanned aerial vehicles, etc. These works have military applications, although they are not specifically developed for the military field;
- Reviews on generic ML and optimization techniques with military applications. In this category, we included several more or less generic review papers on ML and optimization techniques, but always including applications to the military field;

- Reviews on ML and optimization techniques focused on military applications. This category includes works specifically developed for the military field, and therefore can be considered as being more related to the proposals of this article.

Due to its interest, we will analyze this last group in more detail. Firstly, we have found recent review works on specific optimization techniques, such as dynamic programming and its application to the military field [34].

We have also found a review on AI and its applications in the military field [35]. This review describes three main military applications:

- Military AI overview. The main AI projects and milestones carried out by the Defense Advanced Research Projects Agency (DARPA) are examined. In June 2016, the Alpha AI air combat simulation opponent pilot developed by the U.S. University of Cincinnati scored a victory over famed Air Force tactical instructor Colonel Gene Lee. As of 2018, the AI Next project included five directions: new AI capability; robust AI; anti-AI; high-performance AI and next-generation AI;
- Observation, orientation, decision and action approaches. The authors explore this approach based primarily on data and AI in the military field:
    - Information fusion. In the military field, merging heterogeneous information from different sources is essential;
    - Situation awareness. The perception of the existing elements in the volume of the time and space of an environment, understanding its meaning, intention with other agents and its future;
    - Decision support system. Requires the participation of hybrid decision makers: humans and computers;
    - Path planning. Used to avoid threats or obstacles to help commanders choose the appropriate path;
    - Human–machine interface. Interrelating a mechanical element with an assistant in an ideal environment.
- Challenges and solutions of AI in a military context. The main ones found by the authors are as follows:
    - Modeling of complex systems. A large amount of information is generated in a war environment. The level of intelligence can be verified and evaluated by means of a simulation system, which is quite reduced;
    - Imperfect information environment. In a combat situation, the information obtained is always limited, and the authenticity of the information is not guaranteed. AI technology is not omnipotent; it must be combined with traditional technologies, such as knowledge reasoning and search and solution, in which the role of domain knowledge is indispensable.

This work, while interesting, is not specifically oriented to ML, and also focuses almost exclusively on U.S. applications. In this respect, it cannot be considered a global survey of the field.

There are other, less important NATO (North Atlantic Treaty Organization) papers that are coincident with the military applications of AI [31].

It can be seen that the existing scientific literature does not reflect specific and comprehensive studies on ML applied to the military environment. Therefore, we see the need to carry out this novel study, which provides the scientific community with updated data on the scientific interest of ML applied to the military environment, explaining the identified areas and making a categorization of them, and showing the existing interest in decision making based on data in such a military environment. In Section 3 we explain the methodology used for this study, which aims to be more objective than most of the related works by using a bibliometric study as a basis.

*2.2. Background of Data-Driven Architecture for Organizations*

Data-driven decision making has ML algorithms as key components. There are works [37] that specify conceptual architectures that allow an organization to adopt this philosophy in a practical way in a scalable context, which allows it to adapt to a big data environment, i.e., with an increasing data volume and with a variety of formats (unstructured, semi-structured and structured) that are produced; therefore, they have to be processed at high velocity. We have not found in the literature specific architectures for military organizations: therefore, we have specified in Figure 1 a generic architecture, based on [37,38], that could support these types of organizations in the big data context discussed. As mentioned, the main objective of this paper will be to specify in more detail the components of such an architecture for a military organization.

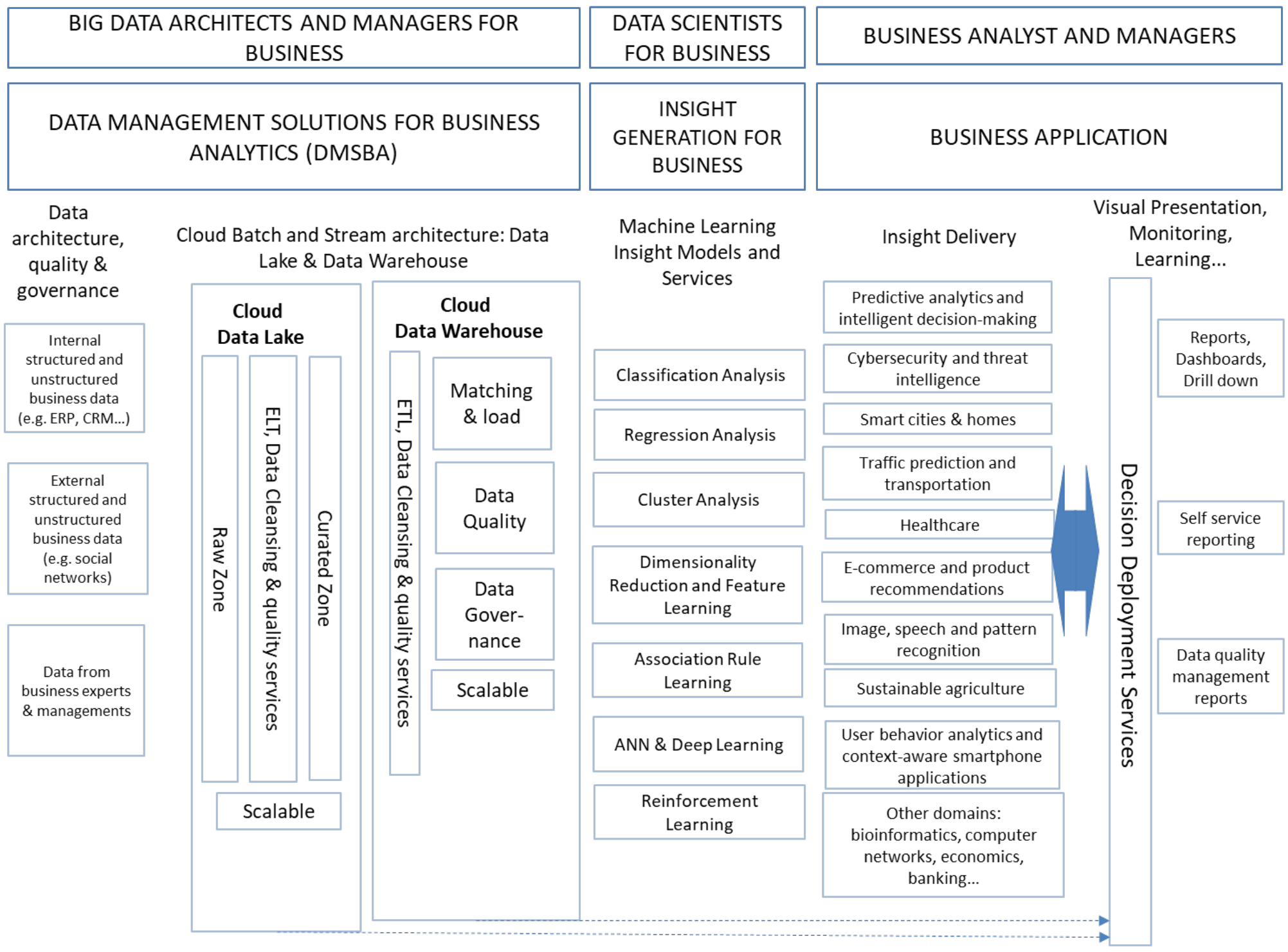

**Figure 1.** Data-driven architecture for nonmilitary organizations.

The components of the conceptual architecture are as follows:

- Data management solutions for business analytics (DMSBA). Data are the essential raw materials for data-driven decision making. From them, ML algorithms will be able to find the desired knowledge in the form of patterns. In this way, we should have access to both internal and external data considered important for decision making. The format of these data can be structured (typically tables of relational databases that are manipulated with SQL), semi-structured (with structure but not tables, including NoSQL databases) or unstructured (without a defined format, such as natural language, images, video, etc.) [37,39]. A conventional organization is mainly characterized by structured internal data from its operational systems: ERP (enterprise resource planning) and CRM (customer relational management). Prior to their analysis, these data must be stored. For this purpose, there are two main types of systems: a data warehouse, mainly for structured data, and a data lake, for all other cases. Data provided by experts should also be considered in this layer, since this type of knowledge should be systematized in organizations [40];
- Insight generation for business. This is the task performed by data scientists with ML algorithms at its core. ML algorithms are classified into three main categories: supervised learning, unsupervised learning and reinforcement learning. Supervised

ML uses labeled data in a certain target class, which must be learned by the algorithm. If this target variable is continuous, we would be in a case of regression, as would be the case if it is discrete in classification. Unsupervised ML works with unlabeled data by obtaining groupings of the data, association rules, dimensionality reduction, etc. Reinforcement ML considers the problem of a computational agent learning to make decisions by a trial-and-error method [39];

- Business Application. This is the task performed by data scientists with ML algorithms at its core. Using the knowledge extracted in the previous layer, the appropriate business decisions are made in this layer. Although there are many business applications, the most important ones have been identified in [38], as shown in Figure 1. A preliminary view is that many of these applications could be applicable in military fields, such as predictive analytics and intelligent decision making, cybersecurity and threat intelligence, healthcare, image, speech and pattern recognition, etc. In this layer there is a constant task of monitoring and permanent learning about the decisions taken, in order to contrast them with the business objectives.

## 3. Research Methodology

In this Section we explain the methodology used, which we will later apply in Section 4. The methodology used is based on the scheme presented in Figure 2.

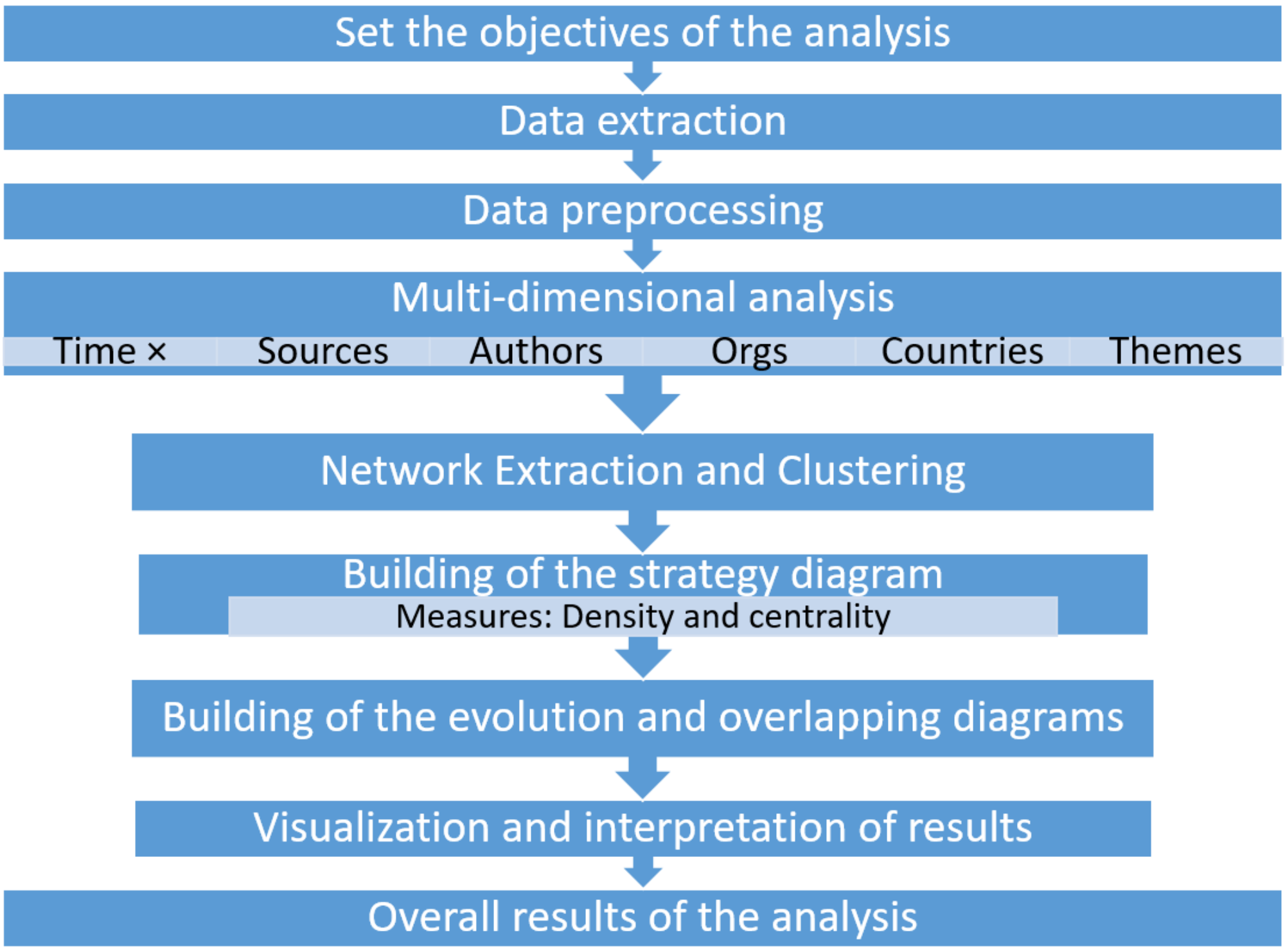

**Figure 2.** Steps used in the research methodology.

### *3.1. Set the Objectives of the Analysis*

At this stage, the research questions should be clearly formulated.

The research topic, the period to be investigated, the sources to be used, etc., must be delimited.

The objective is to conduct a comprehensive study of the relationship between the military and ML, following on from the proposal of conceptual military architecture.

We want to know what is being researched, by whom, in which countries and organizations and how it is developing over time. We want to draw conclusions about where specific research topics could be focused. Finally, our objective is to present the main areas identified that relate ML to the military environment, to group these areas into categories that we will explain one by one in Section 4 and, finally, to comment on the dimensions detected in relation to these topics.

### 3.2. Data Extraction

To carry out this study, we need to know the scientific literature related to the topic "Machine Learning" and, at the same time, relating to the "military*" environment. We have chosen as a timeline all the existing literature up to the year 2021, this year included. Unfortunately, we have not had access to military databases or documents, and we have only used the Web of Science core collection database, which is widely accepted in the scientific environment. The query used in January 2022 was:

TS = ("machine learning" and "military*")

Once we have the data, we proceed to its preprocessing.

### 3.3. Data Preprocessing

From the results obtained in the extraction, we selected those that are really related to ML in the military environment, finally obtaining 525 documents.

Specifically, a standardization process was carried out by merging the plural and singular forms and converting the acronyms into their respective keywords using Levenshtein distance in SciMAT.

### 3.4. Multidimensional Analysis

Qualitative data can be analyzed dynamically using multidimensional analysis techniques [40]. In this way, we will identify the dimensions and the type of analysis used on them, as indicated in Table 2.

**Table 2.** Dimensions, meaning and type of analysis available.

| Dimension | Meaning | Type of Analysis Available |
|---|---|---|
| Sources | Journals, proceedings… | Citation, Bibliographic coupling, Co-citation |
| Authors | Author and co-authors | Co-authorship, Citation, Bibliographic coupling, Co-citation |
| Organizations | Authors' institutions | Co-authorship, Citation, Bibliographic coupling |
| Countries | Authors' countries | Co-authorship, Citation, Bibliographic coupling |
| Themes | Keywords, terms, topics… | Co-occurrence |

Moreover, in this type of study, we can add temporal dimensions. Depending on the type of analysis, we use the following bibliographic relations:

- **Co-authorship analysis**: The relatedness of items is determined based on their number of co-authored documents;
- **Citation analysis**: The relatedness of items is determined based on the number of times they cite each other;
- **Bibliographic coupling analysis**: The relatedness of items is determined based on the number of references they share;
- **Co-citation analysis**: The relatedness of items is determined based on the number of times they are cited together;
- **Co-occurrence analysis**: The relatedness of keywords is determined on the number of documents in which they occur together. In this sense, the equivalence index is usually used [41]:

$$e_{ij} = \frac{c_{ij}^2}{c_i \cdot c_j} \quad (1)$$

where $c_{ij}$ is the number of documents in which keywords $i$ and $j$ co-occur; $c_i$ and $c_j$ represent the number of documents in which each appears.

### 3.5. Network Extraction and Clustering

Based on the aforementioned relationship measures (Table 2) different networks are constructed, depending on the type of analysis. After this construction, a process of clustering or grouping of the items that are considered similar is usually carried out. Therefore, we cluster those nodes that are sufficiently close to each other and sufficiently separated from the rest of the clusters.

In this work for co-occurrence analysis, we use the single-link hierarchical clustering algorithm Agnes, with a network size between 3 and 12. This algorithm [42] is an agglomerative clustering algorithm, i.e., it considers at the beginning that each item is a cluster in itself, and, in each step, it tries to group the nearest clusters or items. Using the single-link option, the distance between two groups is the distance between the closest individuals in each group. Other options, such as complete-link, have been discarded as they tend to generate very large clusters and do not allow the identification of certain thematic interesting areas.

### 3.6. Building of the Strategy Diagram

The strategy diagram, Figure 3, can help to better profile the importance of each cluster in the co-occurrence analysis. It is based on two measures: centrality and density.

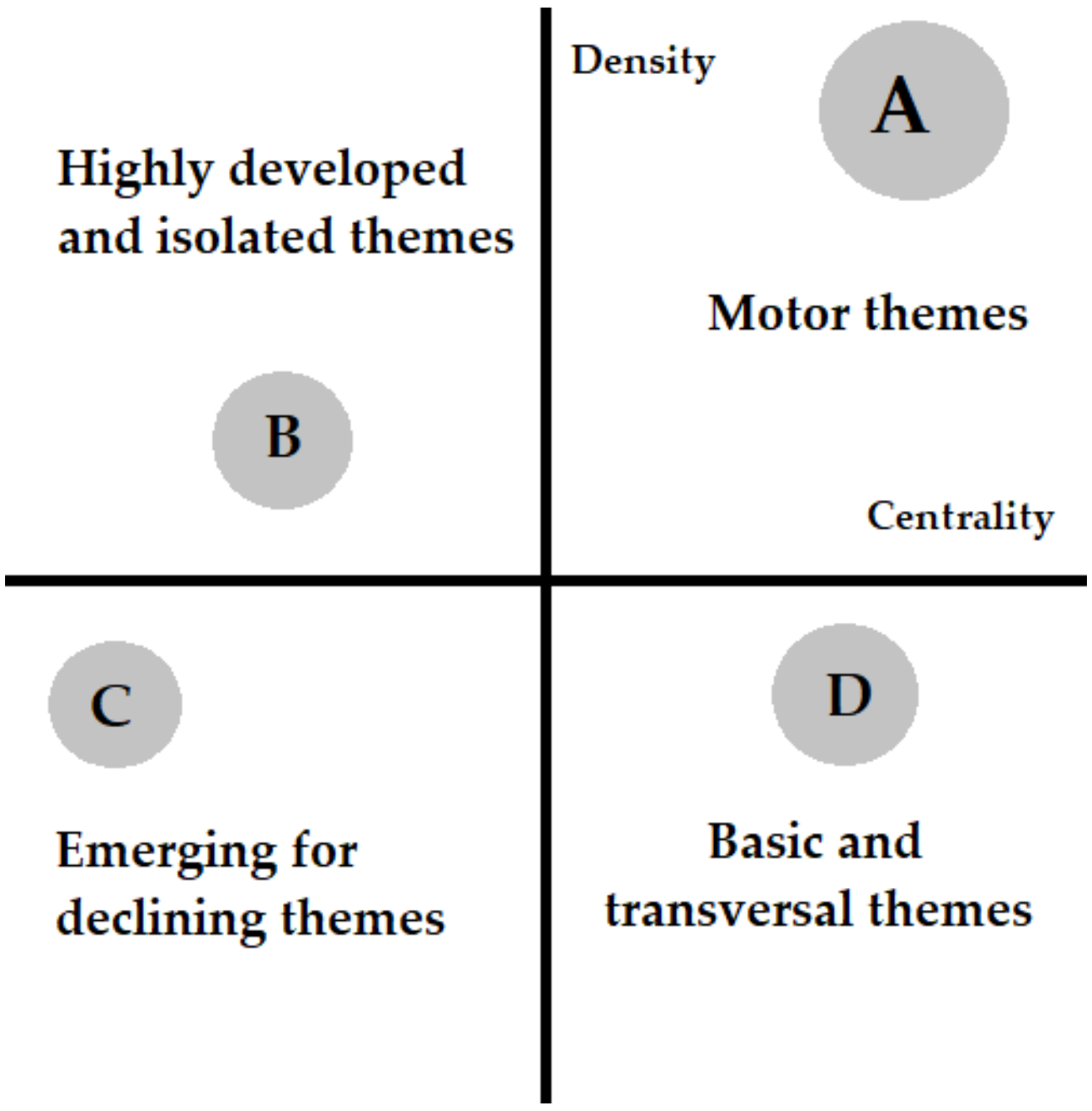

**Figure 3.** Strategy diagram.

The object of centrality will be to measure how the networks relate to the other networks. This value can be understood as a measure of the importance of an item in the development of the entire research field analyzed. It is defined as follows [41]:

$$c = 10 \cdot \sum e_{kh} \quad (2)$$

where $e_{kh}$ has been defined in Equation (1); $k$ and $h$ are keywords, where $k$ is related to the main cluster and $h$ to other clusters.

The density can assess the internal strength of the network or of the item. This value can be considered as a measure of the degree of the development of the item. It is defined as follows:

$$d = 100 \cdot \frac{\sum e_{ij}}{w} \quad (3)$$

where $e_{kh}$ has been defined in Equation (1); $i$ and $j$ are member elements of the set; and $w$ indicates the number of such elements within the group.

Once these measures have been calculated for each cluster, they are presented in a strategic diagram that classifies the themes into four groups: highly developed and isolated themes; emerging or declining themes; basic and transversal themes; and motor themes.

*3.7. Building of the Evolution and Overlapping Diagrams*

We will divide the analysis into two periods: before 2016, and that year plus those years after it. In this way we obtain the evolution and overlapping diagram.

For the evolution diagram we use the Inclusion Index, and for the overlapping map we use the Jaccard Index.

In Figures 4 and 5 we can see the evolution diagram where the Inclusion Index, widely used in financial analysis, is applied [43]. Figures 3 and 4 are based on the many examples that exist [5] about it versus others [44]. In Figure 3 there are two different evolution zones separated by a line. In one is cluster A1 and cluster A2, and in the other are clusters B1, B2 and C2. The solid lines mean that the linked clusters share the main element. The dotted line means that the themes share nonmain elements. The size of the borders indicates the Inclusion Index, and the size of the spheres represents the number of publications associated with the cluster.

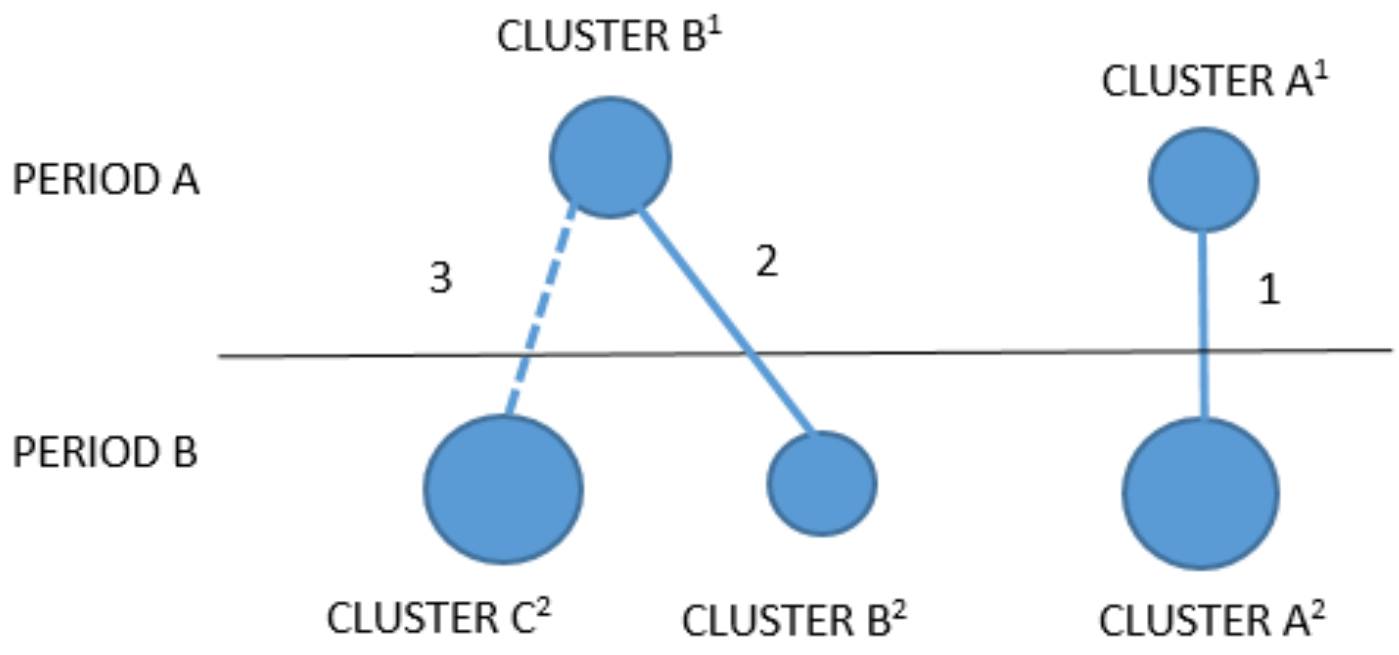

**Figure 4.** Example of evolution areas.

Figure 4 shows the stability measures between consecutive periods (or how much they overlap). The circles are the periods, and the numbers inside are the keywords for each period. The horizontal arrow is the keywords shared between two consecutive periods, and the number in parentheses is the similarity index. The down arrow is the number of outgoing keywords, and the up arrow is the number of incoming keywords in the period.

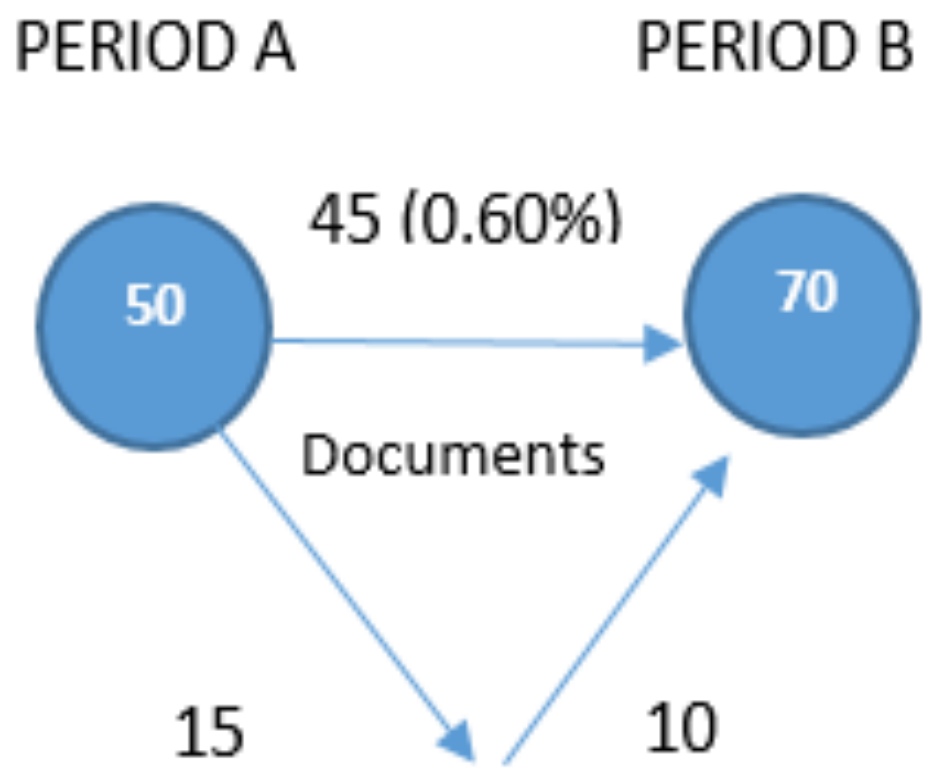

**Figure 5.** Example of stability between periods.

### 3.8. Visualization and Interpretation of Results

This section is responsible for displaying the results obtained, allowing the user to process and evaluate them. By being able to perform these actions on the results, the user focuses on the most important points of the dimensions and analyzes them in detail.

In this work we used SciMAT, VOSviewer and Excel.

### 3.9. Overall Results of the Analysis

Taking into account all the sub-analyses carried out, an attempt is made to respond to the objectives set out in the initial phase.

## 4. Application of ML in the Military Context

Using the methodology described above, the results presented below are obtained. First, we will present the different areas and how we have categorized them, after which we will offer the dimensions that emerged.

### 4.1. Theme Analysis

We present a preliminary analysis by topics in which the co-occurrence has been applied, such that we can see, in Figure 6, the major themes related to ML. It is interesting to see the relationship that links the military environment in relation to ML with areas such as AI, deep learning, algorithm and neural networks, and at the other extreme, but related to the military area, with veterans, depression or the army. These areas will be discussed later.

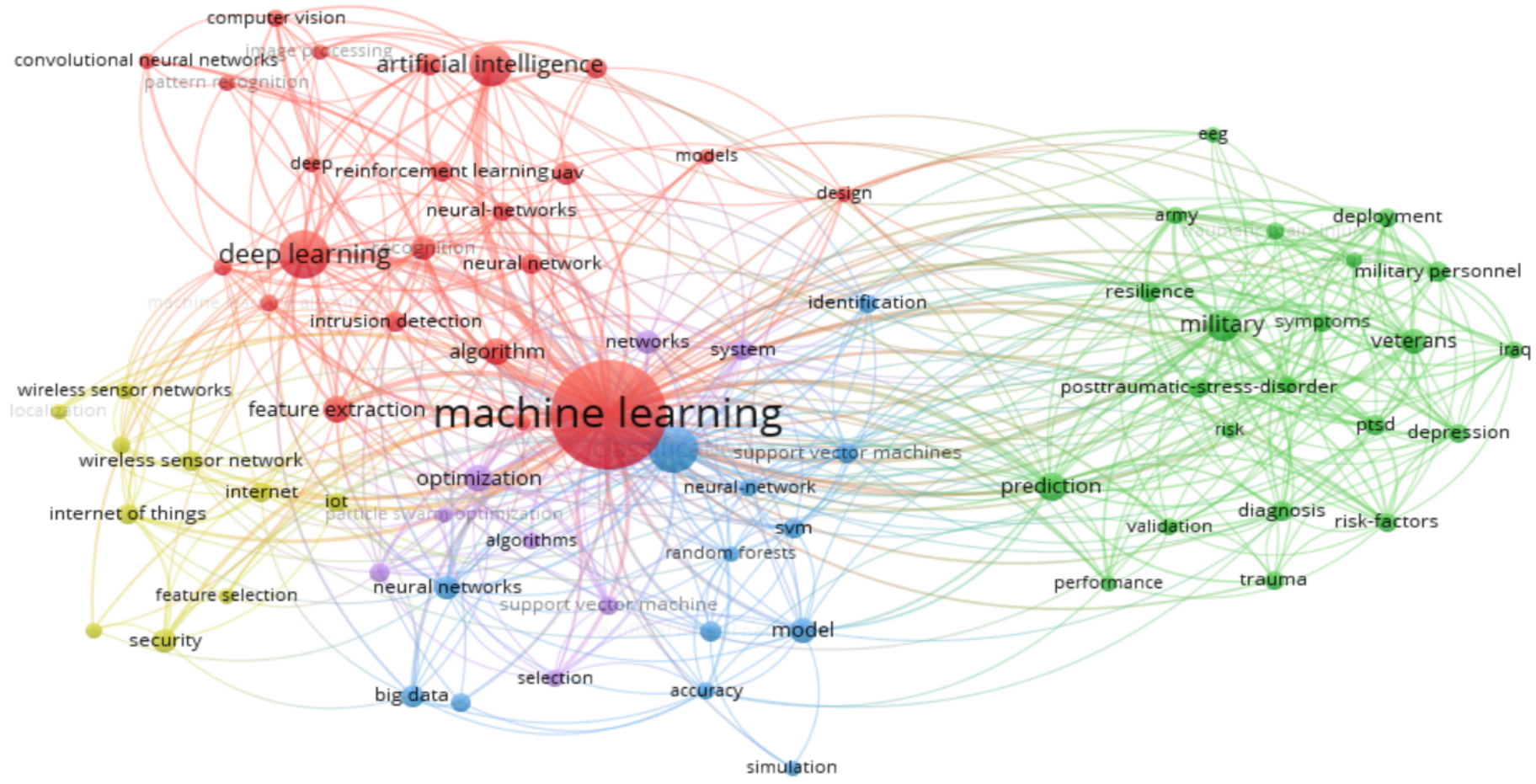

**Figure 6.** Preliminary network by co-occurrence.

Figure 7 establishes a clear distinction between two periods; the one prior to 2016 where publications related to ML and the military environment were limited in scope. On the other hand, since 2016, the number of publications has increased until 2021, coinciding with the pandemic. However, the number of citations on this topic continues to rise, which shows that it is a topic of scientific interest. As can be seen in the Section for period one, there is hardly any distinction between the areas; however, in period two, many interesting areas emerge.

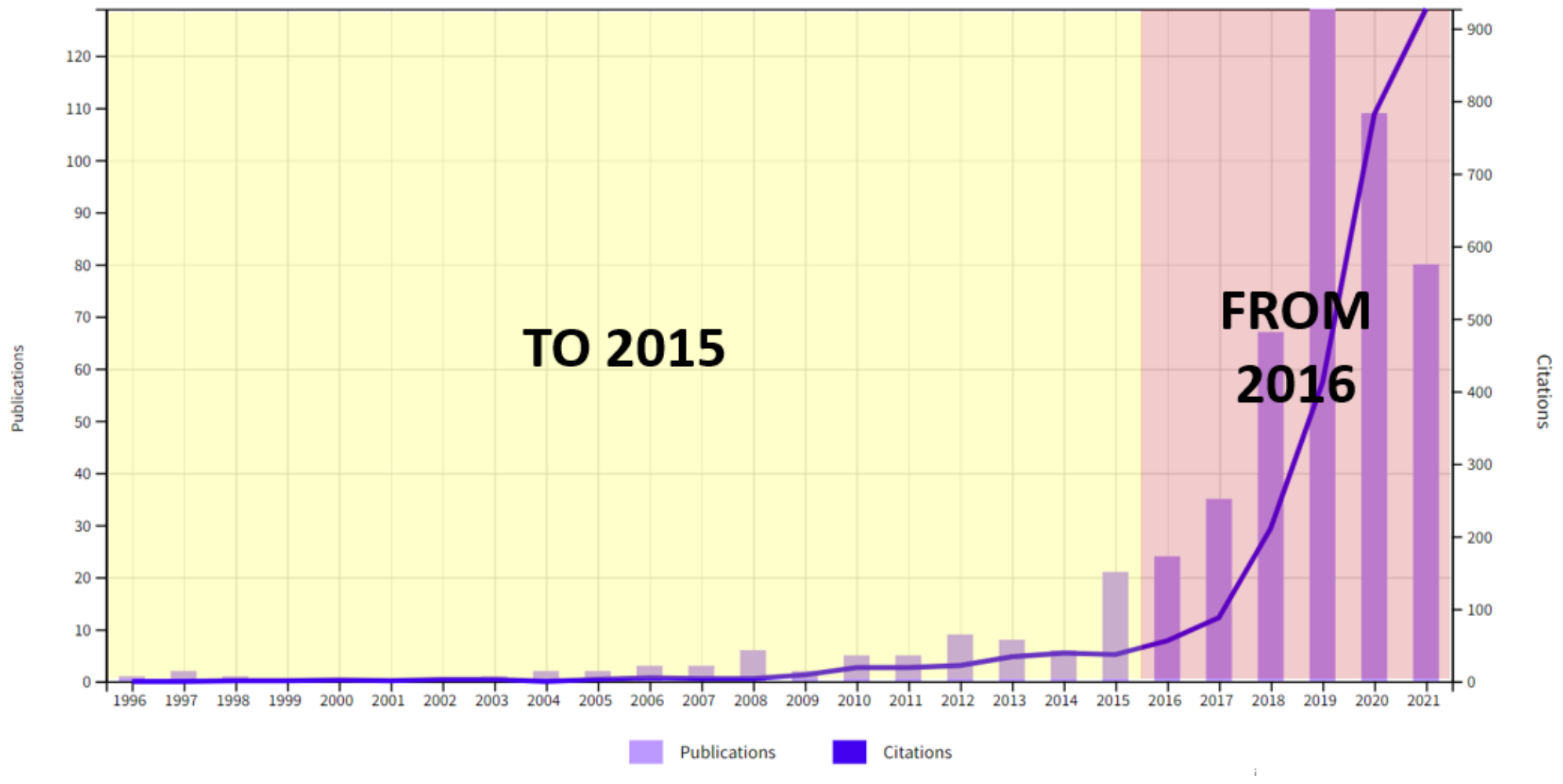
TO 2015
FROM 2016
Publications
Citations
1996 1997 1998 1999 2000 2001 2002 2003 2004 2005 2006 2007 2008 2009 2010 2011 2012 2013 2014 2015 2016 2017 2018 2019 2020 2021
0 10 20 30 40 50 60 70 80 90 100 110 120
0 100 200 300 400 500 600 700 800 900

(**a**)

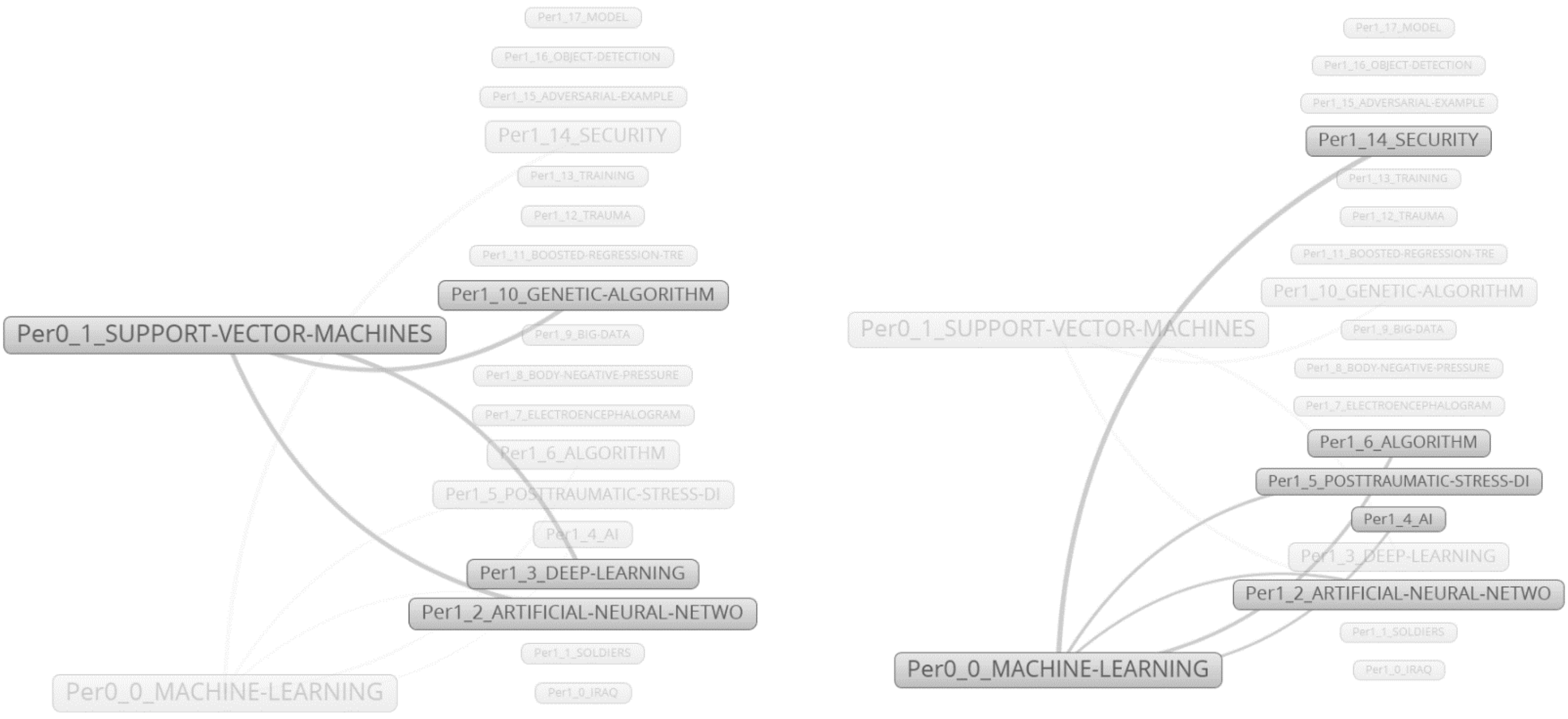

(**b**)

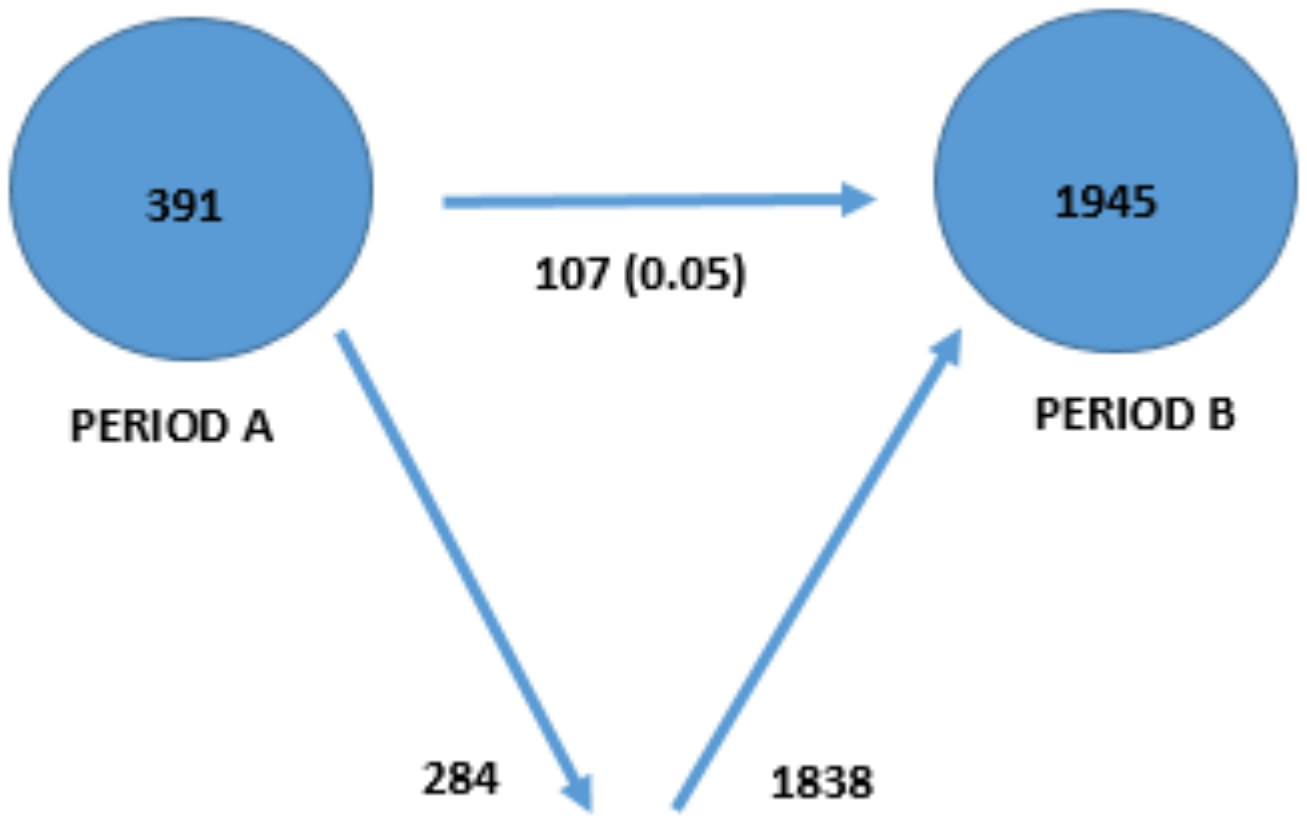

(c)

**Figure 7.** Periods of analysis (**a**) and evolution (**b**) and overlay diagram (**c**).

Period to 2015

Focusing on the area of ML, in Figure 8, and always in relation to the military environment, we can see a network by co-occurrence where different dimensions, such as AI, security or intelligent systems acquire special interest in the 'to 2015' period.

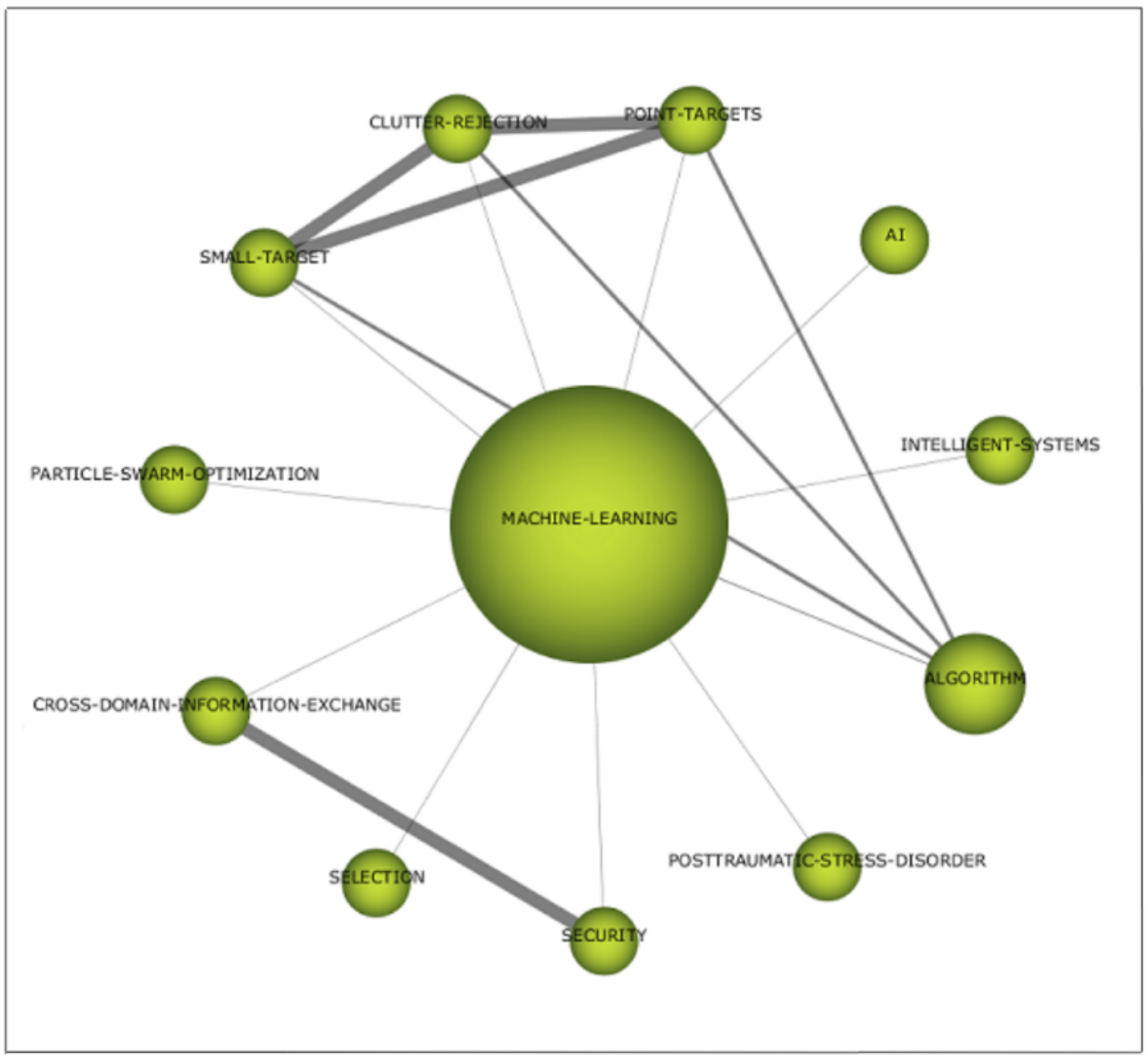

**Figure 8.** Network by co-occurrence; ML; Period to 2015.

Period from 2016

In the period since 2016, more categories have emerged around ML applied to the military world. As we can see in Figure 9, there are different areas that can be divided into five categories.

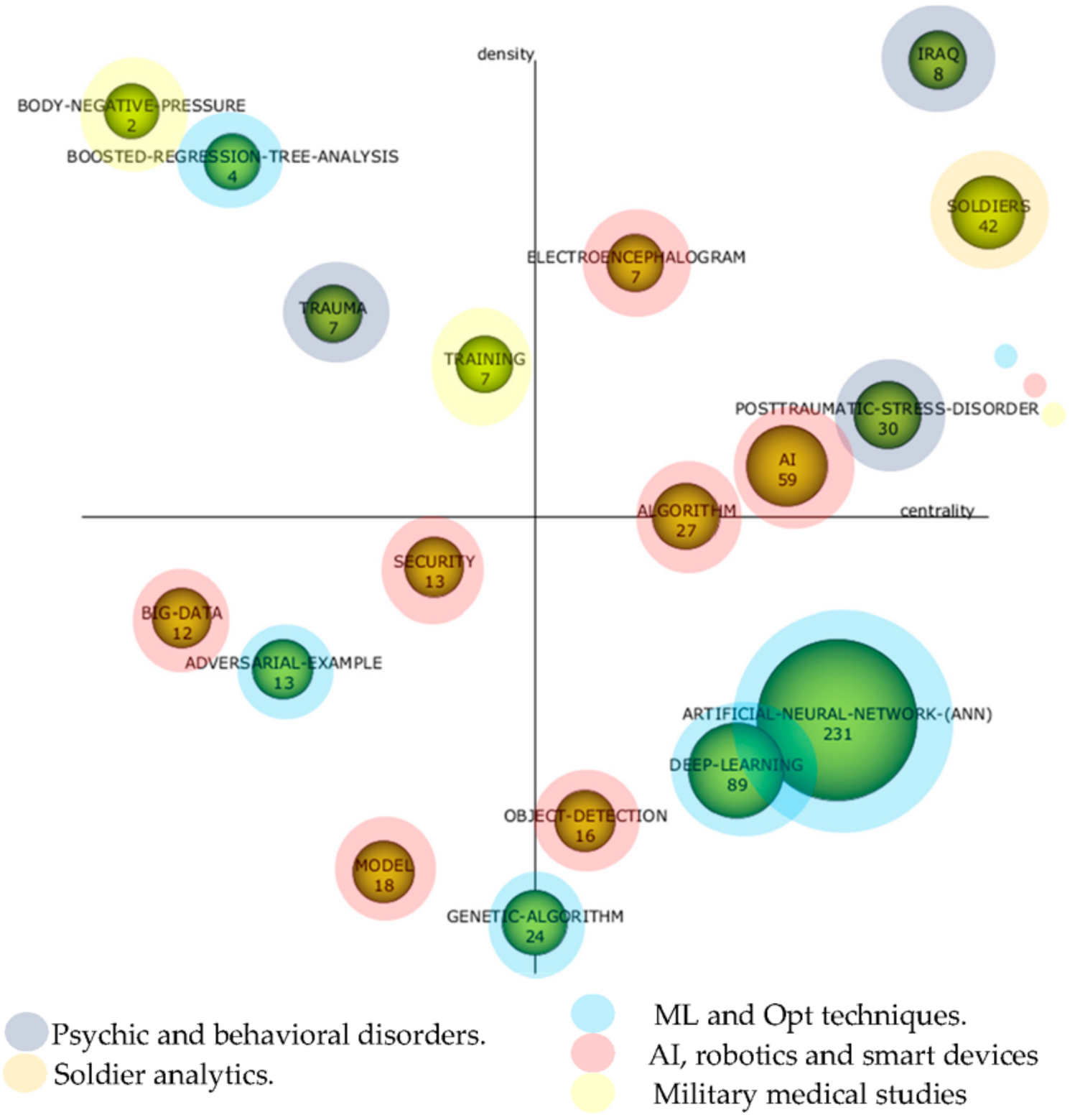

**Figure 9.** Strategy map; Categories from 2016.

Because of their interest, we proceed to study each of these categories in more detail:

Psychological and Behavioral Disorders

In this category we include the themes Iraq, post-traumatic stress disorder and trauma.

We analyze one of the main topics, Iraq, in Figure 10, which we call Iraq and the war in Afghanistan. Within this category, the use of data will address these disorders in relation to the military world.

Psychological disorders: Combat and war conflicts cause stress and mental health problems in civilians, military personnel and veterans, which are analyzed in different studies through literature reviews and data analysis [45]. Behavioral disorders: Through data analysis, studies are conducted to determine which proposed solutions are the most effective for these pathologies related to the military experience [46].

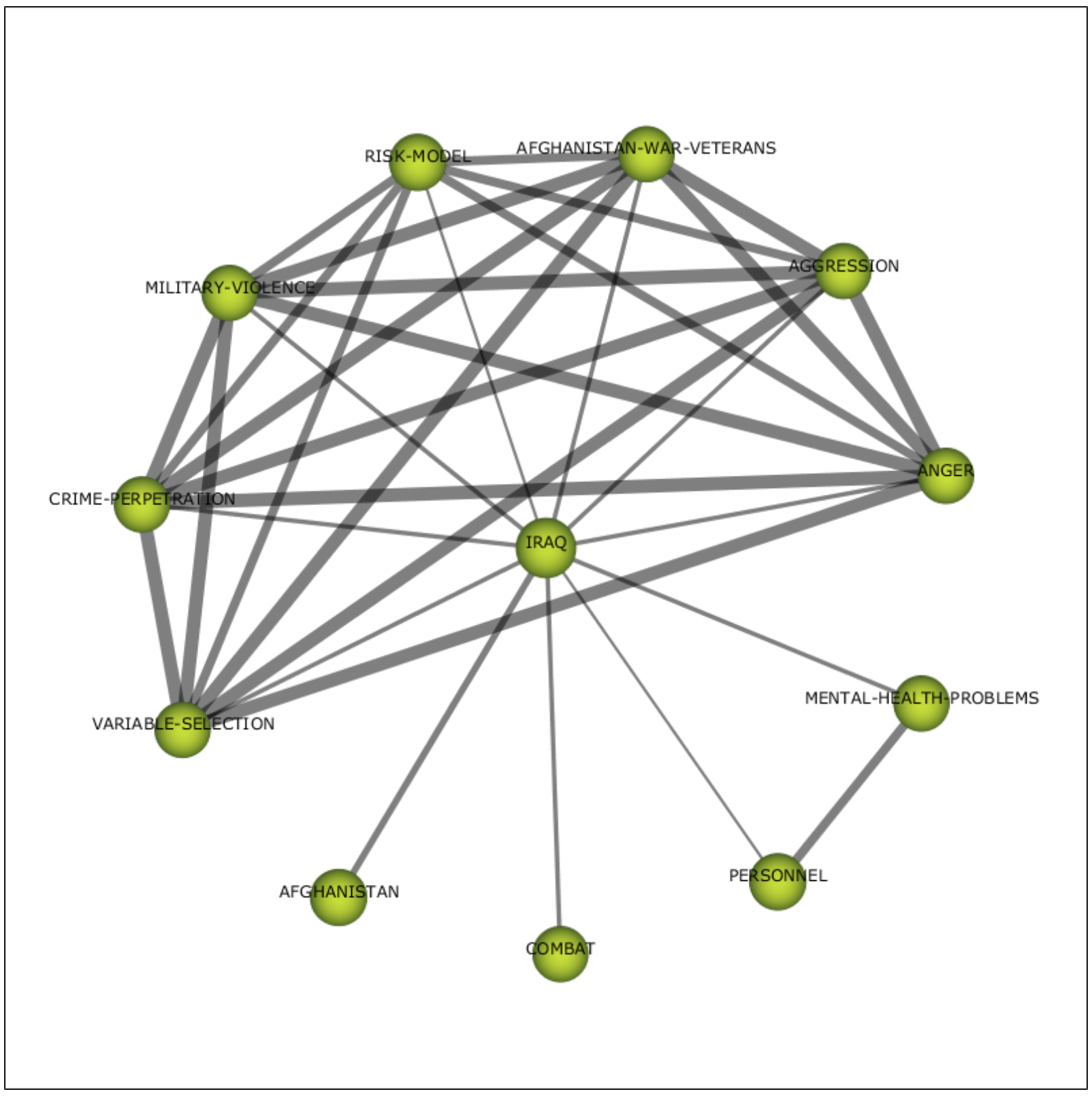

**Figure 10.** Category: Psychological and behavioral disorders; From 2016; Theme: Iraq and Afghanistan wars.

Another major theme within this category is post-traumatic stress disorder, a frequent symptom in veterans [47]. In general, there are studies applying ML to predict this symptomatology, some of these studies in Danish soldiers who participated in Iraq [48] related to psychological disorders. It is also easy to find studies applying analytical studies on prostatic stress in Afghanistan veterans related to conduct disorders [49].

Finally, we discuss trauma in the military environment and how there are studies [50] in which ML is applied. Some trauma data analysis techniques are more focused on mortality [51], while others focus on mental problems, such as shock or stress [52].

Soldier Analytics

In this category we include the *soldier* theme, which in itself has enough weight to be a category.

The *soldier* is a figure on which a large amount of data are analyzed, Figure 11, and whose work is also facilitated by ML [53], providing an advantage, in many cases a strategic advantage, over the enemy, in addition to creating in their environment different areas of study. Applications in these areas, such as resilience, prevention, diagnosis, depression, AI or ML can help to save; currently, there are already studies focused in this direction, such as clinical decision support systems to focused, detailed assessments of suicide risk in patients considered as being at high risk [54].

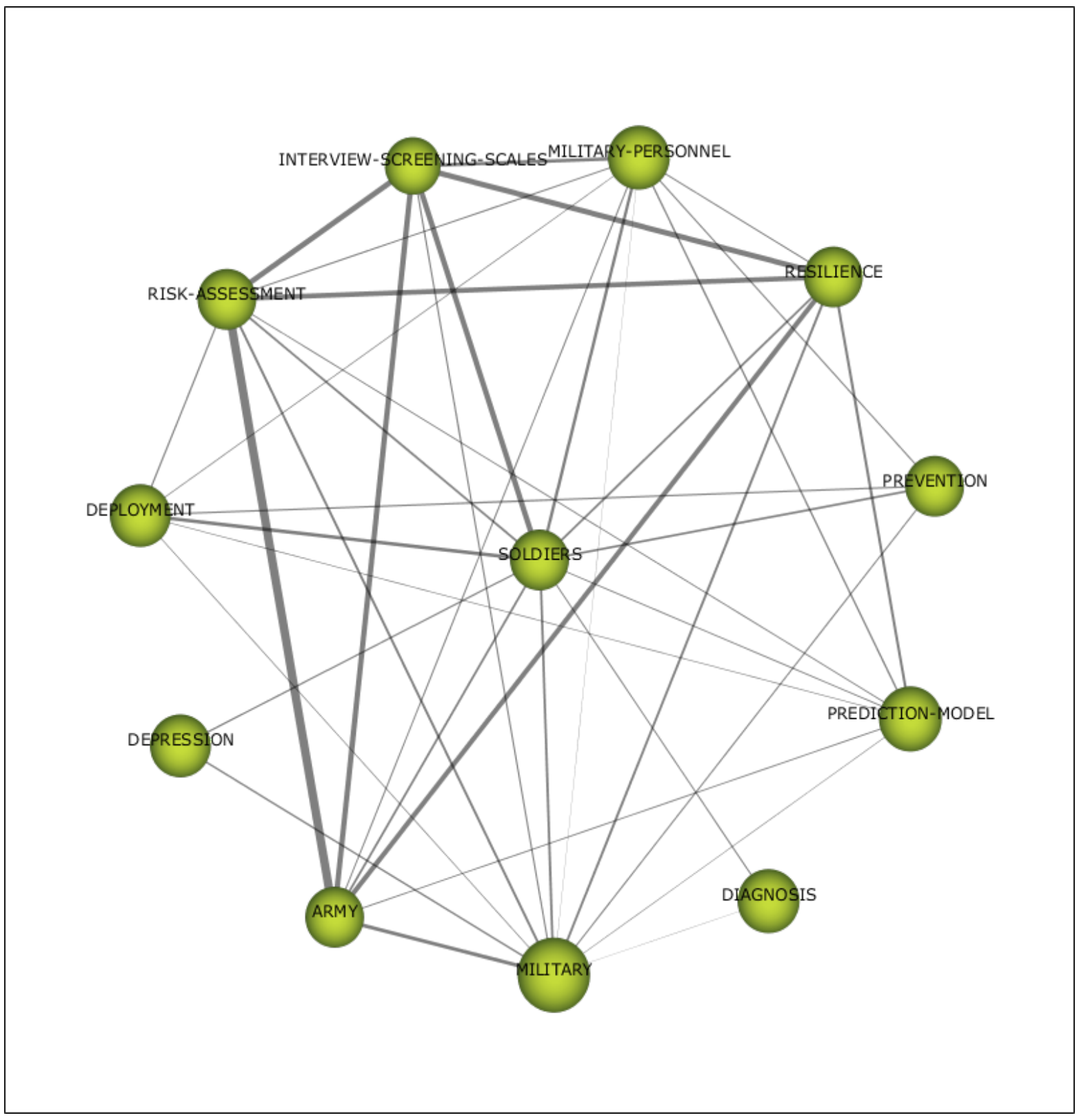

**Figure 11.** Category: Soldier analytics; From 2016; Theme: Soldiers.

ML and Opt Techniques

In this category we include the themes artificial neural network (ANN), deep learning, boosted regression tree analysis, genetic algorithms and adversarial example.

The area of ANN belongs to the ML and Opt techniques' category. Through the area *ANN* and the corresponding data processing, in a transversal way, we collaborate with different areas, such as soldiers or psychological and behavioral disorders; for example, studies were used to highlight the variables that at the beginning of compulsory military service increase the stress of the military by means of a prediction model based on ANN [55].

Within this category, as shown in Figure 12, we have the deep learning area, which is a cross-cutting theme that through its use allows the military world to obtain very useful automatic learning in their daily work, such as identifying any person anywhere and preventing crimes even before they happen [56].

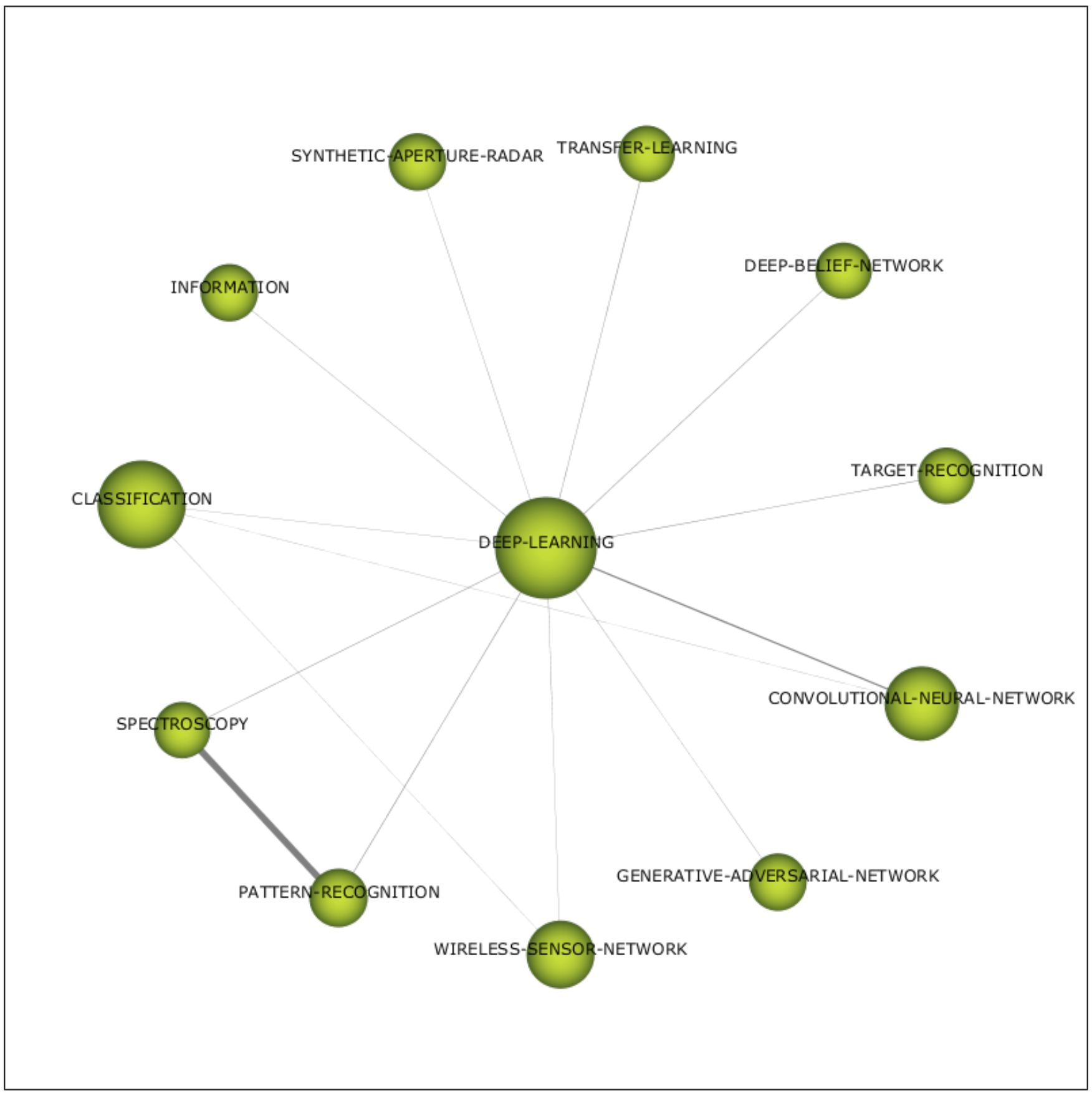

**Figure 12.** Category: ML and Opt techniques; From 2016; Theme: Deep learning.

We see within the same category the boosted regression tree analysis, related to link adaptation and K-nearest neighbors, which belongs to the scenario of highly developed and isolated topics, being a category of little relevance.

The genetic algorithm also belongs to the same category, and we see its relationship with optimization, reinforced learning and clustering algorithms. It can be seen how military applications produce a large amount of data collected in the battlefield, as well as how these data are amenable to processing by genetic algorithms, using crossover and mutation probabilities that are automatically adjusted at each generation [45].

Adversarial example ML is a research area within this category, which focuses on the design of strongly developed ML algorithms in adversarial environments [57].

AI, Robotics and Smart Devices

In this category we include the themes AI, electroencephalogram, security, algorithm, object detection, model and big data.

AI is an area of high demand in the use of data in relation to the military world, as seen in this category, Figure 13. This is largely due to the increase in military investment in AI research advances [58]. It highlights the use of drones and the IoT, to such an extent that military drones in some cases are mixed with commercial ones [59], and how AI influences military strategy [60].

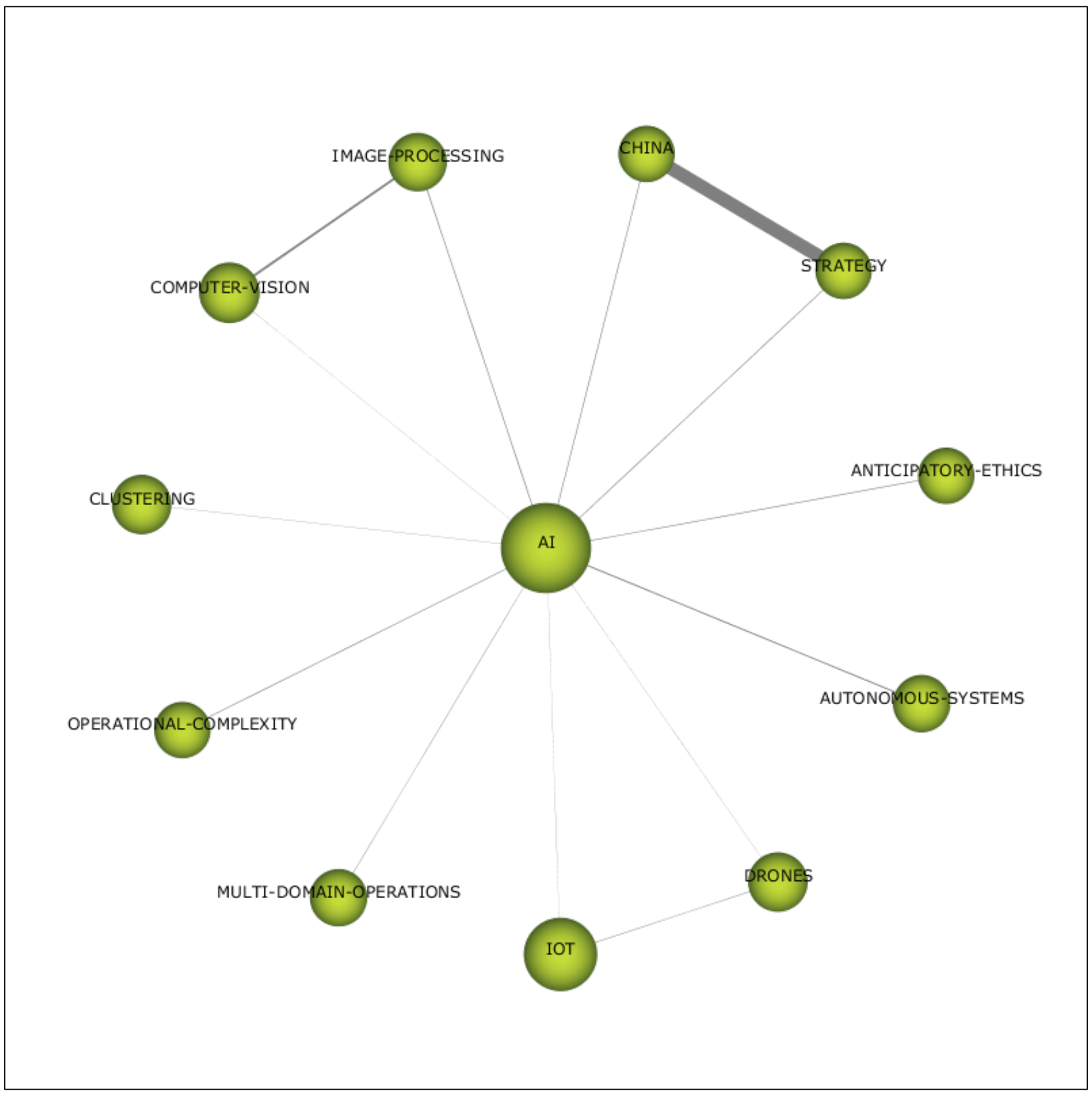

**Figure 13.** Category: AI, robotics and smart devices; From 2016; Theme: Artificial intelligence.

The electroencephalogram is related to data processing [61], although its presence in the military world is still of little significance.

ML algorithms collaborate with cybersecurity and, in turn, can be affected by cyberattacks, so there are more and more studies on them [62]. Technological *security* and protecting data are vital in the military environment.

The need for algorithms in ML related to the military world is a fact and is present in the interpretation of the data obtained [63].

Unmanned aerial vehicles are a reality and include object detection by means of algorithms; some complex ones are also using deep learning [64].

The *model* theme, supported primarily by design, is an emerging area that is beginning to have value with other areas, such as trauma [51], within this category.

The military experience provides a large amount of data, *big data*, which need to be analyzed by ML and is already being completed in many cases for medical purposes [65]. We can see how this area relates to data mining in the military environment in scientific studies [66].

Military Medical Studies

In this category, we include the themes body negative pressure and training.

In relation to body negative pressure, we found evidence from algorithm-supported studies on military data that focus on military medical studies [67].

The models used in military *training* are based on data processing and are becoming increasingly common in military training [68].

### 4.2. Source Analysis

We can observe in Figure 14 how the number of publications related to ML applied to the military world has clearly been increasing since 2015, although during the pandemic years, it suffered a slight decline at the same time that citations continued to rise.

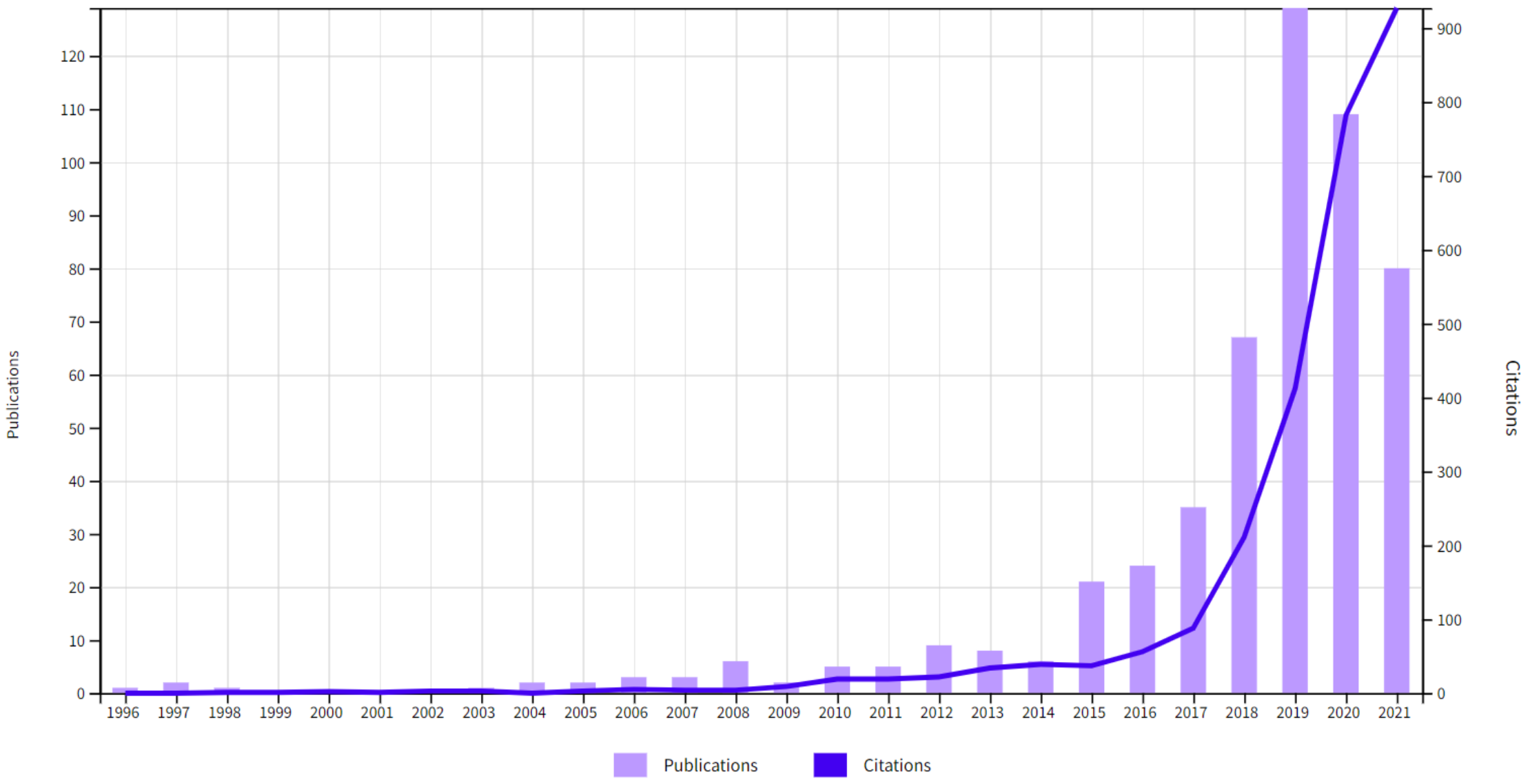

**Figure 14.** Trend and citation structure.

The areas "Electrical-Electronic Engineering", "Computer" and "Telecommunications" are identified as the main categories in which ML relates to the military world; see Table 3.

**Table 3.** Main areas.

| Web of Science Categories | Record Count | % of 525 |
|---|---|---|
| Engineering Electrical Electronic | 161 | 30.667 |
| Computer Science Artificial Intelligence | 119 | 22.667 |
| Computer Science Information Systems | 92 | 17.524 |
| Computer Science Theory Methods | 77 | 14.667 |
| Telecommunications | 77 | 14.667 |
| Optics | 63 | 12.000 |
| Computer Science Interdisciplinary Applications | 34 | 6.476 |
| Automation Control Systems | 24 | 4.571 |
| Remote Sensing | 23 | 4.381 |
| Instruments Instrumentation | 21 | 4.000 |
| Imaging Science Photographic Technology | 20 | 3.810 |
| Engineering Multidisciplinary | 19 | 3.619 |
| Psychiatry | 19 | 3.619 |
| Computer Science Software Engineering | 18 | 3.429 |
| Physics Applied | 17 | 3.238 |
| Chemistry Analytical | 16 | 3.048 |
| Computer Science Cybernetics | 16 | 3.048 |
| Computer Science Hardware Architecture | 16 | 3.048 |
| Engineering Aerospace | 15 | 2.857 |
| Operations Research Management Science | 13 | 2.476 |
| Robotics | 13 | 2.476 |
| Environmental Sciences | 12 | 2.286 |
| Materials Science Multidisciplinary | 11 | 2.095 |
| Neurosciences | 11 | 2.095 |
| Surgery | 11 | 2.095 |

### 4.3. Country Analysis

Once the number of citations and documents published on ML oriented to the military world by country is normalized, we proceed to their comparison, as we can see in Figure 15. We can observe that, in some cases, the number of publications and citations do correspond, as in the case of China, but in other cases, such as England, Israel and Iran, the number of citations is much higher than the number of publications, which means that the quality of these publications is very high. The opposite is the case in India, South Korea and Canada, where the number of citations is much lower than the number of publications, meaning low quality.

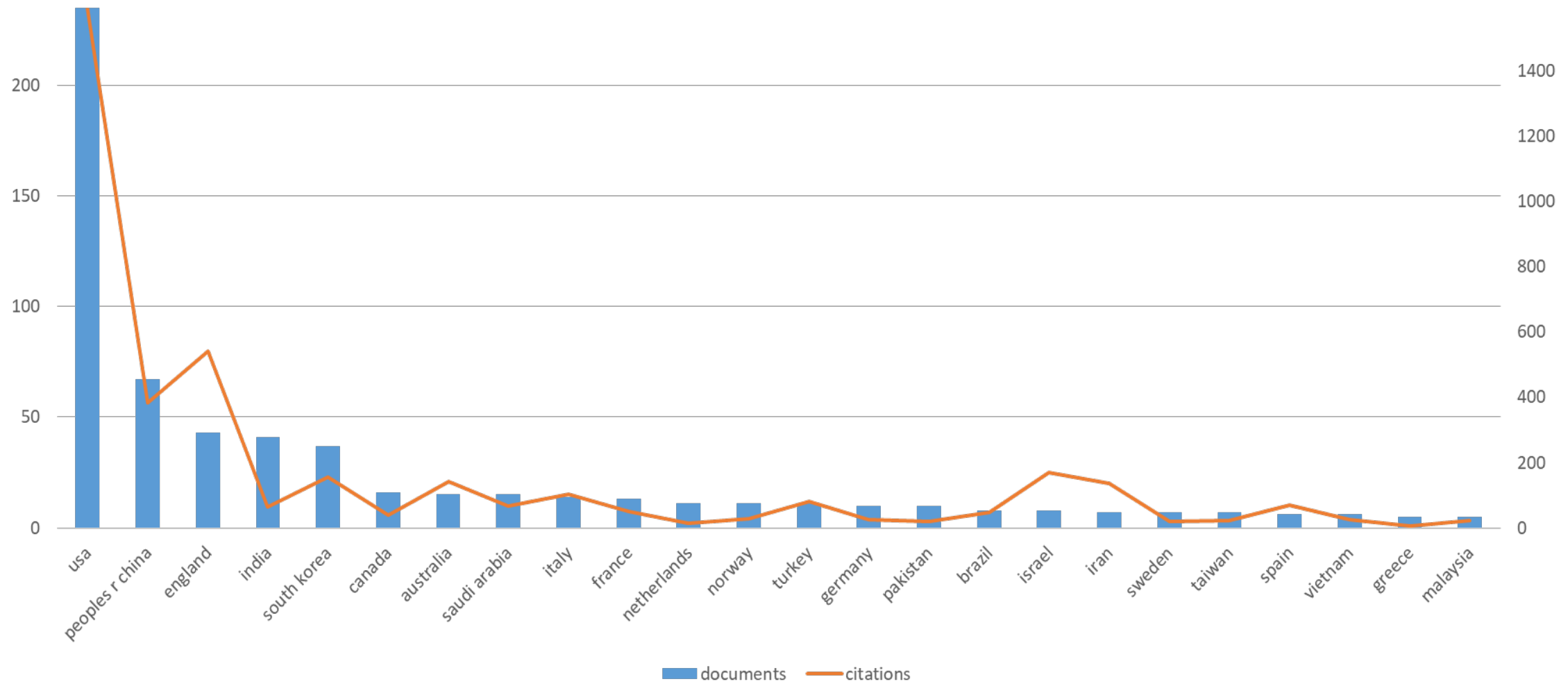

**Figure 15.** Country analysis; Most important by documents and citations.

In Figure 16, we can see in general terms the relationship between different countries in terms of joint participation in the writing of scientific papers on ML applied to the world of work, and therefore the relationship of the co-authorship of researchers from different countries in the writing of these papers. The size of each circuit is proportional to the number of papers, and subsequently we will identify the main countries in the production of scientific material in this topic during the period of highest productivity: 2016–2019. The USA co-author network has been changing in recent years, starting in 2016 with a collaboration network mostly with France and the Netherlands, followed in 2017 and 2018 mainly with England, Australia, Greece and continuing with the Netherlands, until finally in 2019 the main relationship is with China, India and Taiwan. We can see how England centered in 2016 its network of co-authors in France and the Netherlands, collaborating in 2017 and 2018 with the USA, Australia, Israel, Iran and Turkey to finally establish stronger co-authorship with China, India or Saudi Arabia, starting in 2019. China shares co-authorship in ML scientific publications in the military world with Norway in 2016, evolving with the USA, Australia and England during 2017 and 2018 until 2019, when the relationship with India, Canada, Saudi Arabia or Turkey is strengthened.

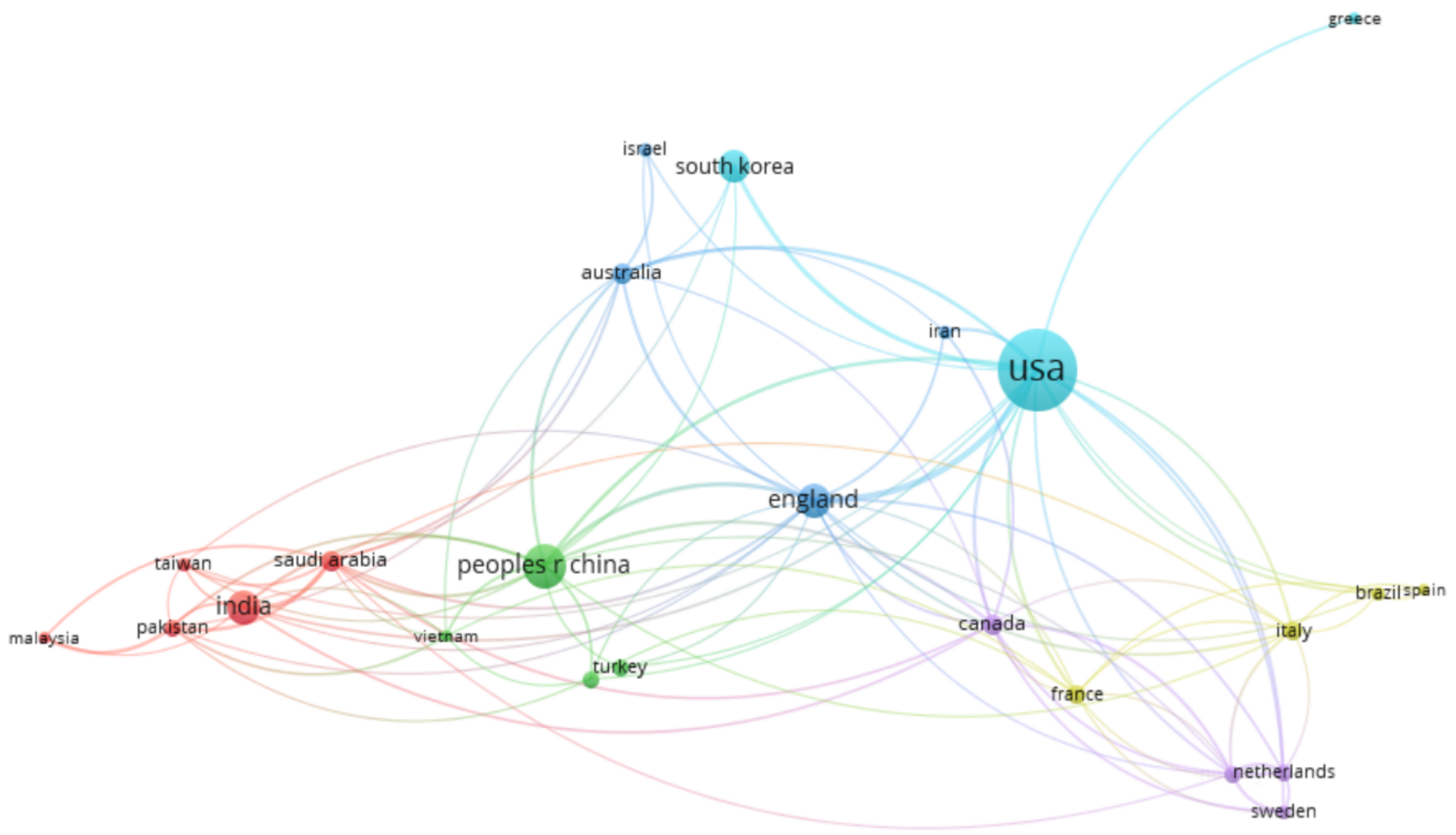

**Figure 16.** Network by co-authorship.

## *4.4. Organization Analysis*

In Figure 17, we normalize by organization the number of scientific papers on ML applied to the military world in relation to different organizations.

It is striking to see how few citations the US army organization has in relation to the number of publications it generates; the same happens with the Korea Advanced Institute of Science and Technology, the Pennsylvania State University, Sejong University, Florida State University, Georgia Institute of Technology and Korea University.

On the other hand, we can see the high number of citations of organizations that have published to a lesser extent, which means that these publications are of high quality, such as the Uniformed Services University of the Health Sciences, Harvard Medical School, the University of Pittsburgh, Boston University, VA San Diego Healthcare System, the University of California San Diego and the University of Michigan.

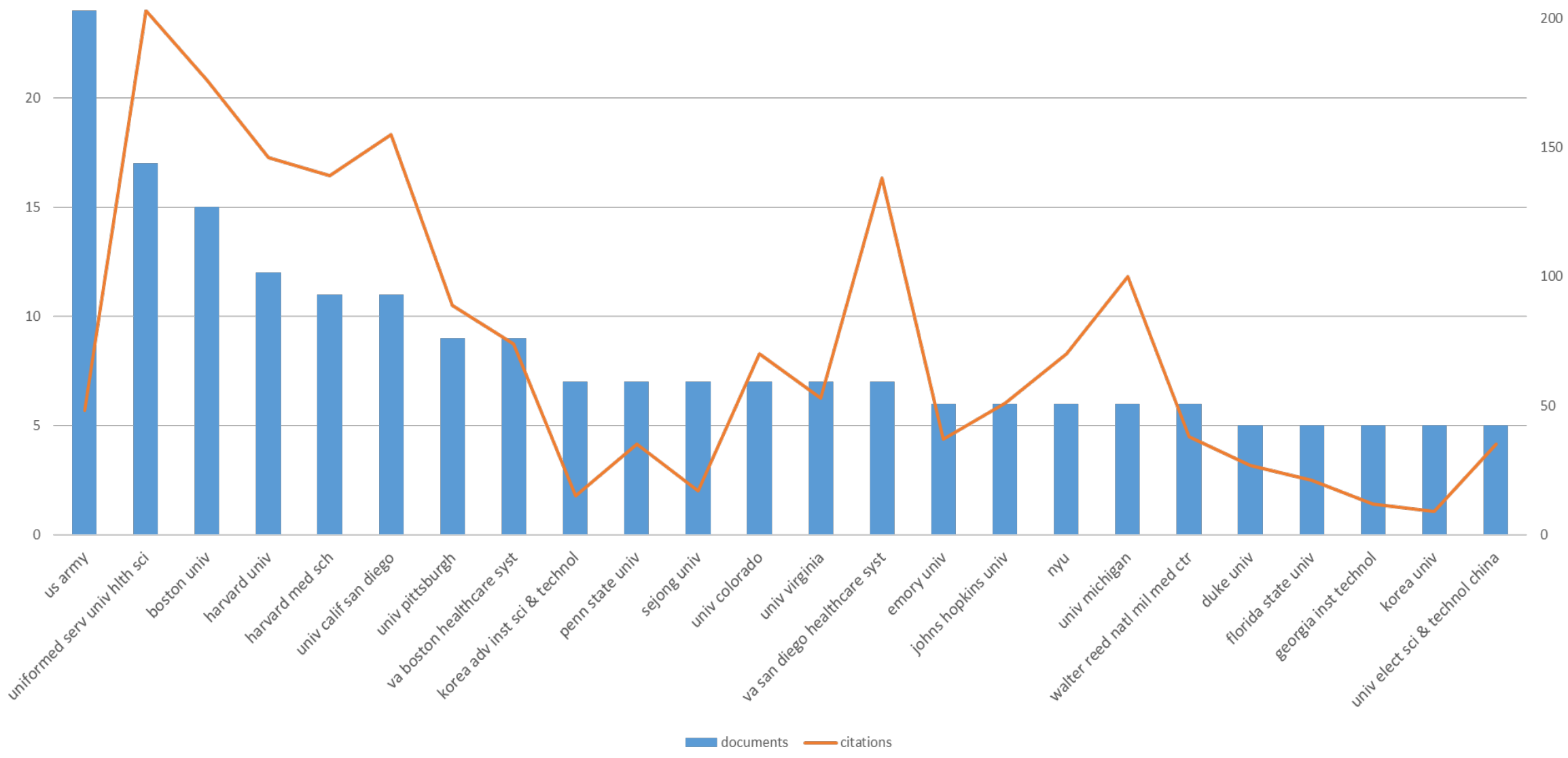

**Figure 17.** Org analysis; Most important by documents and citations.

We present in Figure 18 the co-authorship network between the different organizations in the most significant period, between 2018 and 2019. First, we present where the most significant organizations are: The US Army establishes a relationship in 2018 with the Uniformed Services University of the Health Sciences, the University of California San Diego and the University of Virginia to generate scientific articles on ML applied to the military world. This relationship continues with the University of Colorado and Harvard University, ending with Emory University in 2019; the Uniformed Services University of the Health Sciences begins its co-authorship in 2018 with several organizations, but, before long, in 2019 it focuses its relationships primarily on the Johns Hopkins University, in this case being their only relationship by mid-2019 (Figure 19).

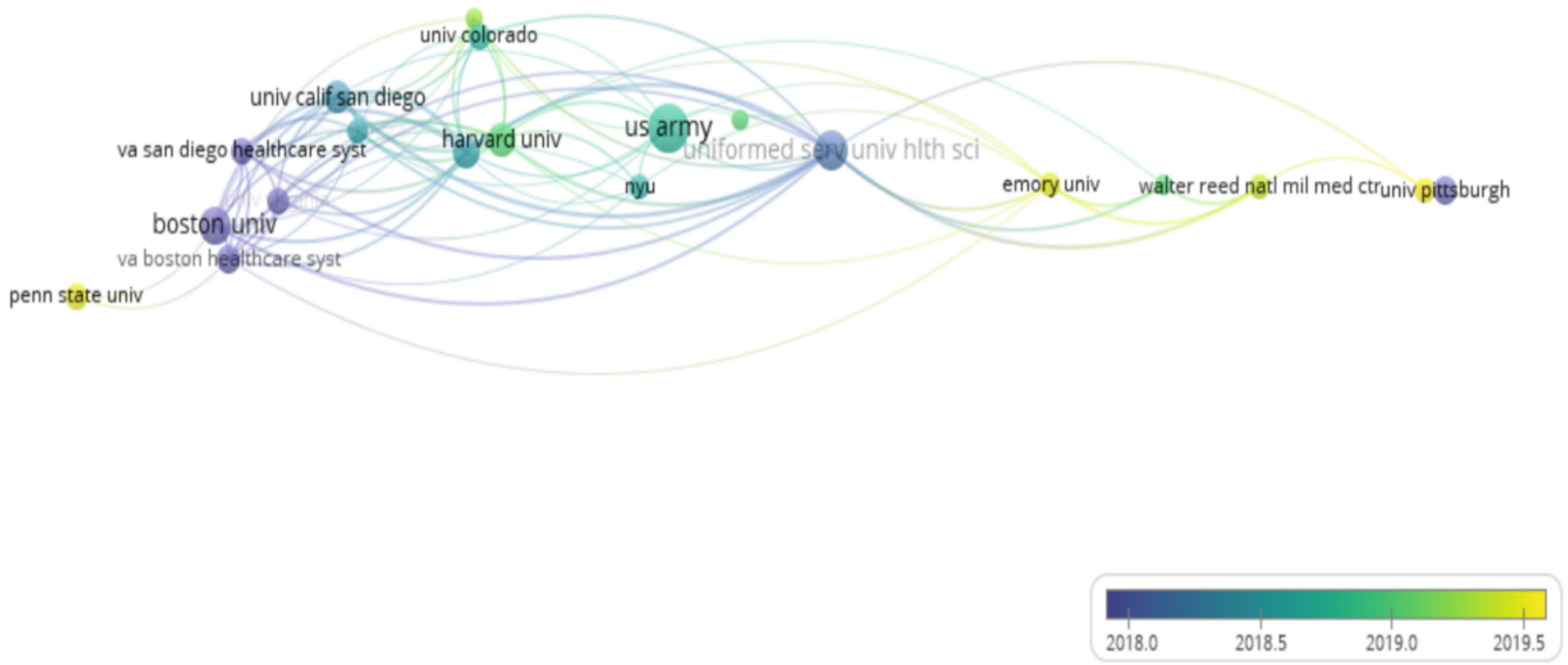

**Figure 18.** Network by co-authorship.

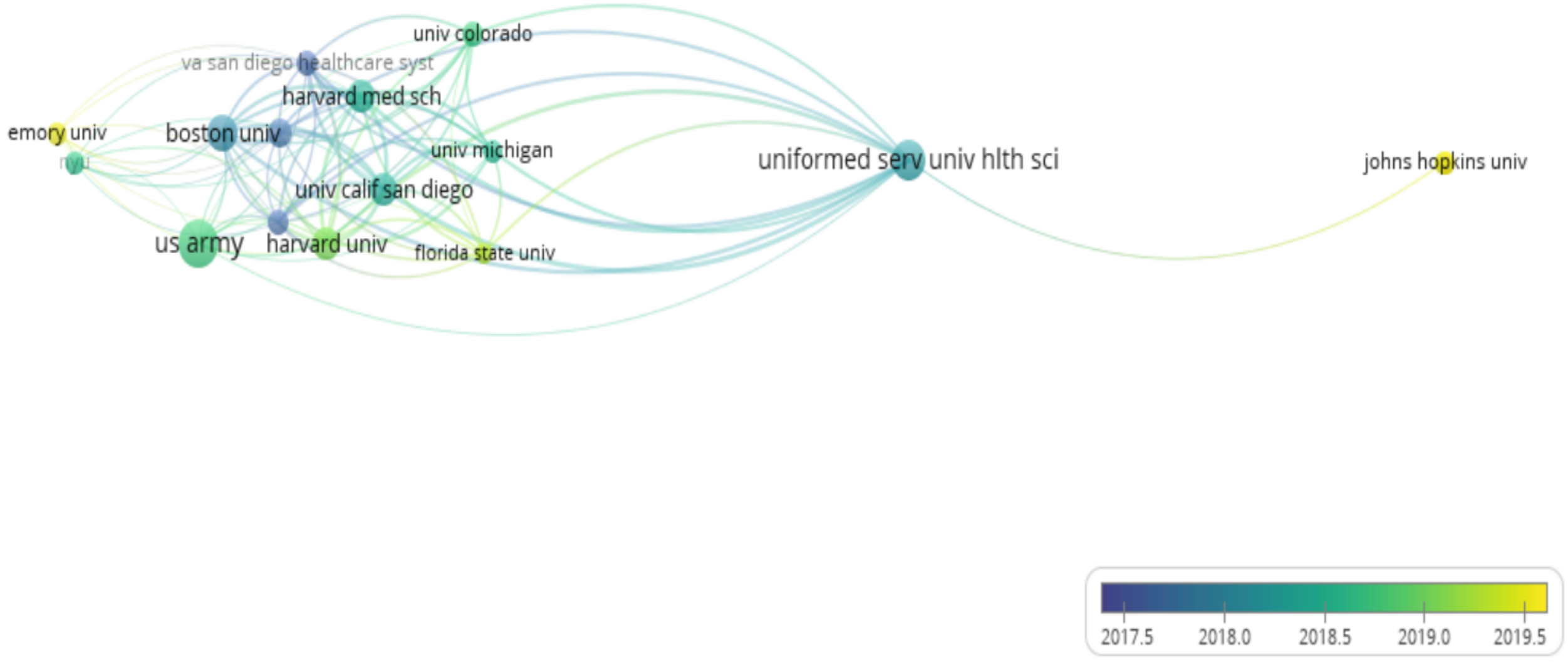

**Figure 19.** Network by citation.

Bibliographic collaboration between different organizations was very strong at the beginning of 2018, and a decrease is seen at the beginning of 2019, as can be seen in Figure 20.

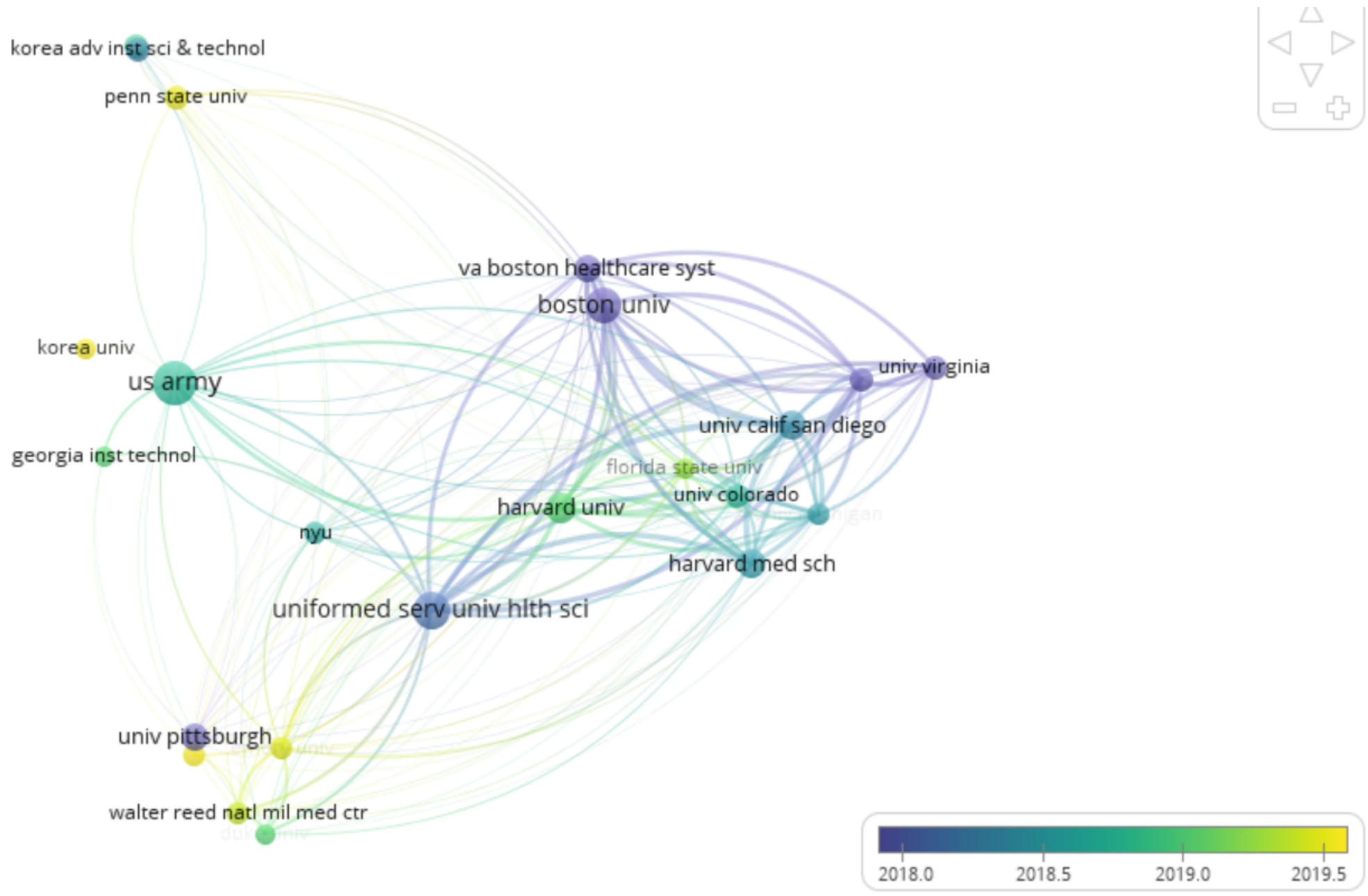

**Figure 20.** Network by bibliographic coupling.

*4.5. Author Analysis*

In an analysis by the authors, we recognize the most relevant authors by their number of publications, citations and year, highlighting, in 2018, Ben-David for the highest number of citations, as we can see in Figure 21. Arie Ben-David works at the Holon Institute of Technology, Holon, Israel, and has more than 24 publications and more than 850 citations to their name.

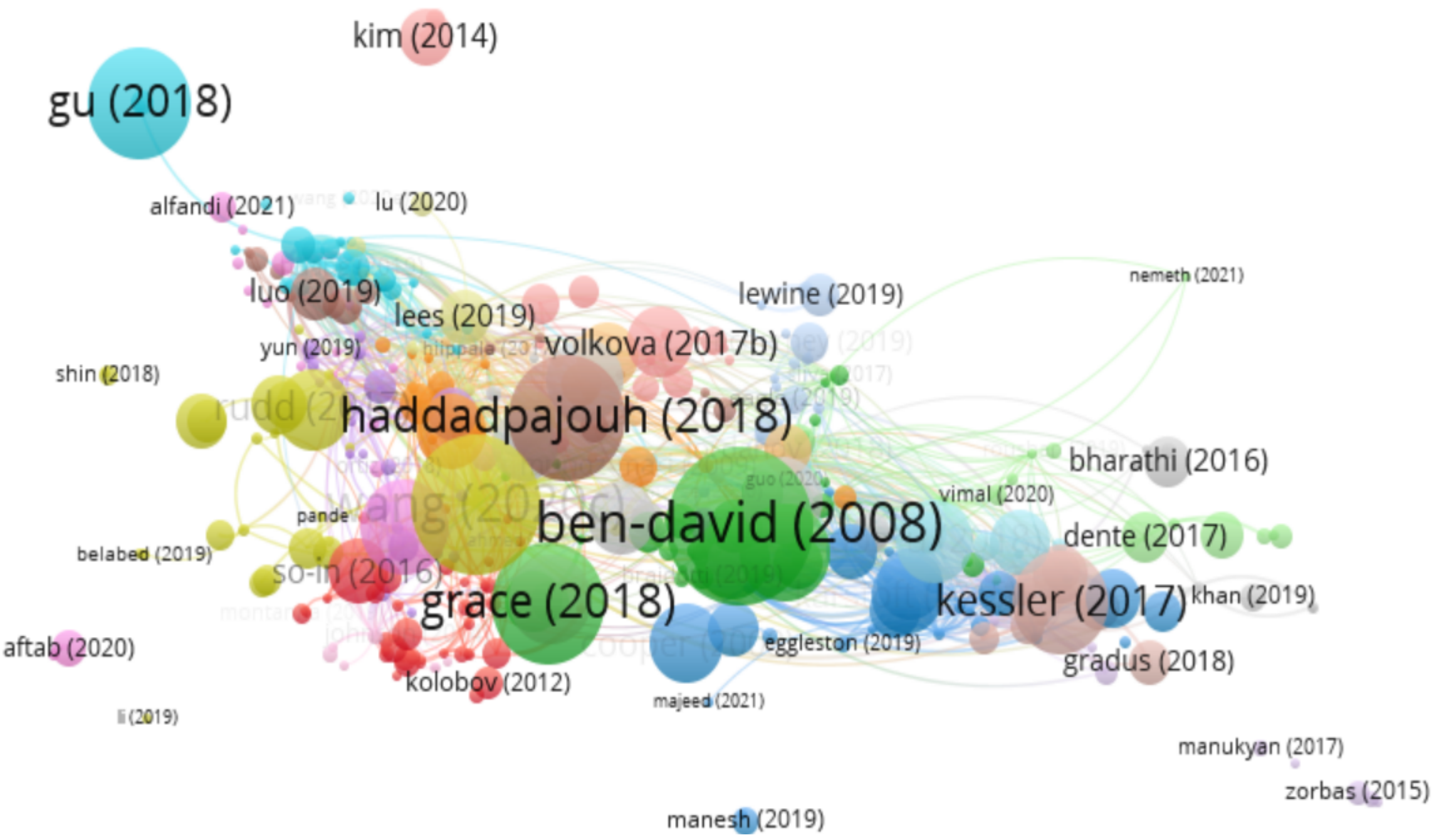

**Figure 21.** Network by citations.

### 4.6. Overall Results of the Analysis

The exponential evolution of the number of publications possibly slowed down from 2020 due to the effects of COVID-19. The most interesting analysis corresponds to the period since 2016. There is a predominance of journals on engineering, electricals, electronics and computing sciences [69].

The USA, China, the UK and South Korea lead in publications and citations. There is a predominance of US universities and the US Army in publications and citations. The main thematic categories have been identified and their importance and perspectives have been characterized:

- Psychological and behavioral disorders;
- Soldier analytics;
- ML and Opt techniques;
- AI, robotics and smart devices;
- Military medical studies.

## 5. Discussion

Based on the bibliometric analysis performed, in this section, we redefine the conceptual architecture presented in Section 2.2. by adapting it to the military context, as shown in Figure 22.

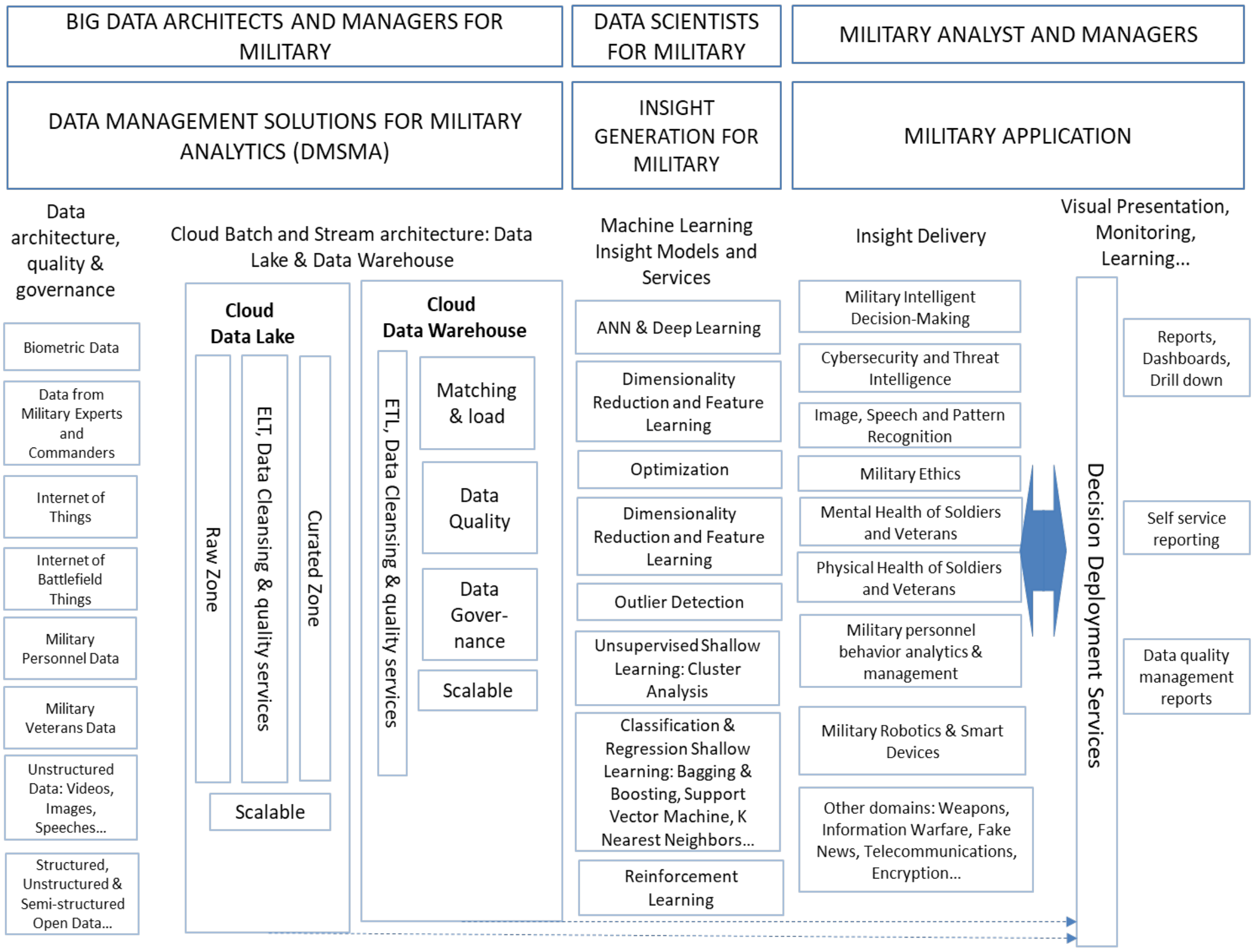

**Figure 22.** Data-driven architecture for military organizations.

The components of each layer are explained below:

- Data management solutions for military analytics (DMSMA). In general terms, it can be said that the data that predominate in the military context are more complex to process than those in a conventional organization. In the study we have carried out, the main sources found are biometric data, e.g., from electroencephalograms [70]; data from military experts and commanders (e.g., field assessments and weaponry needed in that particular field), which should be used in conjunction with the other data to build decision systems [35]; the Internet of things, smart and connected devices widely used by the military, generating large volumes of information over time [71], e.g., those provided by radars [72]; the Internet of battlefield things, which connects soldiers with smart technology in weapons and other objects to give troops “extra sensory powers” [73]; military personnel data, e.g., those obtained in screening interviews [74]; military veterans’ data obtained, e.g., from their administrative personal files [75]; and other data, both internal and open in any format (including videos, images and speeches), e.g., those coming from unmanned aerial vehicles [76], videos for facial recognition [77,78], data from the Internet, etc.;

- Insight generation for military. The predominant ML algorithms in this layer are the so-called deep ML algorithms, since these algorithms give better results with a large volume of data and/or unstructured data [69]. These algorithms are based on neural networks [79]. If the number of layers of such a network is high, they are called deep learning, e.g., these algorithms are applied to military subjects in [72,80,81]. Deep adversarial algorithms are often used to attack other ML models and cause their failure, e.g., [82]. Another deep variant is the long short-term memory (LSTM) algorithms, which are specialized in the treatment of large-scale time series, e.g., [83]. Another type of deep network is the convolutional neural networks (CNNs) that are often used for object detection [82]. In contrast to deep algorithms, there are shallow algorithms [83]; application examples can be found in [84]. We can also find examples of clustering [85], the detection of outliers [86], optimization [87], reinforcement learning [88], etc.;
- Military application. Several of these applications have already been discussed in Section 4.1. The following is a list of the most important ones: cybersecurity and threat intelligence [88]; image, speech and pattern recognition [89]; the mental health of soldiers and veterans [74]; military ethics [90]; military personnel behavior analytics and management [91,92]; military robotics and smart devices [63]; and the physical health of soldiers and veterans, etc. It is remarkable that several of these ML applications in the military field have not been identified in previous works [36].

## 6. Conclusions and Future Work

The objective of this work has been fulfilled: we have carried out a research work presenting initially an ML architecture model applied to a nonmilitary organization, after which a bibliometric study on the use of ML applied to military organizations is presented and, finally, we have applied this study to the original model to obtain an architecture model to apply ML to a military organization. All this has been executed while taking into account that previously there was a lack of scientific information on this subject.

A clear map of the present, past and future of research has been provided.

It has shown a real application of ML and the growing real interest in applying it, in this case to the military field, observing how it is increasingly used to analyze data for automatic decision making.

The following aspects are highlighted:

- The wars in Afghanistan and Iraq are coming under intense scrutiny for their mental effects on military personnel;
- Soldier analytics could become an area in its own right, as it has a lot of specificity compared to today’s people analytics;
- The area of deep learning is growing in military applications;
- There are interesting emerging topics related to AI, such as intrusion detection, brain interfaces, self-driving vehicles, false information processing, cybersecurity, etc.;

- Underlying this is an intrinsic importance in medical research that is likely to have fewer strategic constraints from governments.

We encountered the following limitations: we have not had access to specialized military libraries, and bias is assumed in several publications due to the subject matter.

As lines of future work, the following is highlighted as having a longer development path. The usefulness of ML for the management of military personnel in the style of other already-consolidated areas, such as people analytics [93], has been proven. However, this new and specific area, which we could call soldier analytics, is not yet defined in the literature.

**Author Contributions:** Conceptualization, J.J.G. and R.A.C.; methodology, J.J.G. and R.A.C; software, J.J.G. and R.A.C; validation, J.J.G., R.A.C. and A.L.; formal analysis, J.J.G. and R.A.C; investigation, J.J.G.; resources, J.J.G., R.A.C. and A.L.; data curation, J.J.G. and R.A.C.; writing—original draft preparation, J.J.G.; writing—review and editing, J.J.G., R.A.C. and A.L.; visualization, J.J.G.; supervision, A.L. and R.A.C.; project administration, J.J.G. and R.A.C.. All authors have read and agreed to the published version of the manuscript.

**Funding:** This research has been partially supported support from the FEDER funds provided by the National Spanish project PGC2018-096509-B-I00.

**Institutional Review Board Statement:** Not applicable.

**Informed Consent Statement:** Not applicable

**Acknowledgments:** The authors would like to acknowledge the financial support from the FEDER funds provided by the National Spanish project PGC2018-096509-B-I00 and the DATA SCIENCE AND SOFT COMPUTING FOR SOCIAL ANALYTICS AND DECISION AID Research Group. LaTorre thanks the Spanish Ministry of Science (grant PID2020-113013RB-C22). Carrasco thanks the Spanish Ministry of Science (grant PID2019-103880RB-I00).

**Conflicts of Interest:** The authors declare no conflict of interest.